\documentclass[journal, 11pt,one column,twosides]{IEEEtran} 

\IEEEoverridecommandlockouts

\usepackage{graphicx}
\usepackage{subfigure, caption}
\usepackage{color}
\usepackage{epsfig}
\usepackage{amssymb}
\usepackage{amsmath}
\usepackage{amsthm}
\usepackage{latexsym}
\usepackage{tikz}
\usetikzlibrary{decorations.markings,positioning,arrows,calc,patterns,backgrounds}
\usepackage{pgfplots}
\usepackage{mathrsfs}
\usepackage{bbm}
\usepackage{setspace, hyperref}
\usepackage{epstopdf} 
\usepackage{mathtools}
\usepackage{fancybox}
\usepackage[noadjust]{cite}
\usepackage[hang,flushmargin]{footmisc}
\mathtoolsset{showonlyrefs}

\usepackage{geometry}

\newgeometry{top=25.6mm,bottom=19.5mm,right=19.5mm,left=19.6mm}

\newcommand{\cB}{{\mathcal B}}

\newcommand{\cM}{{\mathcal M}}
\newcommand{\cN}{{\mathcal N}}

\newcommand{\cJ}{{\mathcal J}}

\newcommand{\cP}{{\mathcal P}}

\newcommand\independent{\protect\mathpalette{\protect\independenT}{\perp}}
\def\independenT#1#2{\mathrel{\rlap{$#1#2$}\mkern2mu{#1#2}}}

\newcommand{\ep}{\epsilon}

\newcommand\MC{{ \ - \!\!\circ\!\! - \ }}

\newtheorem{theorem}{Theorem}
\newtheorem{lemma}[theorem]{Lemma}
\newtheorem{proposition}[theorem]{Proposition}

\renewcommand*{\thefootnote}{\fnsymbol{footnote}}
\theoremstyle{remark}
\newtheorem*{remark*}{Remark}
\newtheorem*{remarks*}{Remarks}

\theoremstyle{definition}
\newtheorem{definition}{Definition}

\newtheorem{example}{Example}

\allowdisplaybreaks

\begin{document}

\date{}
\title{ Reconstructing Gaussian sources by spatial sampling}
\author{ Vinay Praneeth Boda$^\dag$ }
\maketitle 

\renewcommand{\figurename}{Figure}

{\renewcommand{\thefootnote}{} \footnotetext{

\noindent 
$^\dag$V.~P. Boda is with the Department of
Electrical and Computer Engineering and the Institute for Systems
Research, University of Maryland, College Park, MD 20742, USA.
E-mail: praneeth@umd.edu.

\noindent This work was supported by the U.S.
National Science Foundation under the Grant CCF-1319799.}}

\thispagestyle{plain}
\pagestyle{plain}

\vspace*{0.1cm}

\begin{abstract} 
 Consider a Gaussian memoryless multiple source with $m$ components 
 with joint probability distribution known only to lie in a given class of 
 distributions.
 A subset of $k \leq m$ components are sampled and compressed 
 with the objective of reconstructing all the $m$ components
 within a specified level of distortion under a mean-squared
 error criterion. In Bayesian and nonBayesian settings, 
 the notion of universal sampling rate distortion function for Gaussian sources is introduced to 
 capture the optimal tradeoffs among 
 sampling, compression rate and distortion level.
 Single-letter characterizations are provided for the universal 
 sampling rate distortion function. Our achievability proofs
 highlight the following structural property: it is optimal
 to compress and reconstruct first the sampled components of the GMMS alone,
 and then form estimates for the unsampled components based on the 
 former.
  
\end{abstract}

\vspace*{0.05cm}

\begin{IEEEkeywords} 
\noindent  Fixed-set sampling, Gaussian memoryless multiple source, 
sampling rate distortion function, universal sampling rate distortion  
function.
\end{IEEEkeywords}

%%%%%%%%%%%%%%%%%%%%%%%%%%%%%%%%%%%%%%%%%%%%%%%%%%%%%%

%%%%%%%%%%%%%%%%%%%%%%%%%%%%%%%%%%%%%%%%%%%%%%%%%%%%%%

\section{Introduction}

Consider a set $\cM$ of $m$ jointly Gaussian memoryless sources with
joint probability density function (pdf) known only to belong to
a given family of pdfs.
A fixed subset of $k\leq m$ sources are sampled at each time instant and 
compressed jointly by a (block) source code, with the objective of reconstructing {\it all}
the $m$ sources within a specified level of distortion under a
mean-squared error criterion.
``Universality'' requires that the sampling and lossy compression code 
be designed without precise knowledge of the underlying pdf.
In this paper we study the   tradeoffs
-- under optimal processing -- among 
sampling, compression rate and distortion level.
This study builds on our prior works \cite{BodNar17, BodNar17U}
on sampling rate distortion for multiple discrete 
sources with {\it known} joint pmf
and universal sampling rate distortion for multiple discrete sources
with joint pmf known only to lie in a {\it finite} class of pmfs,
respectively. Here, we do not assume the class of pdfs to be finite.

\vspace*{0.1cm}

Problems of combined sampling and compression have been
studied extensively in diverse contexts for discrete and Gaussian
sources. Relevant works include lossless compression of analog
sources in an information theoretic setting \cite{WuVer10};
compressed sensing with an allowed detection error rate or quantization
distortion \cite{ReeGas12}; sub-Nyquist temporal sampling of Gaussian sources 
followed by lossy reconstruction \cite{KipnisGold16}; and rate distortion 
function for multiple sources with time-shared sampling \cite{LiuSimErk12}.
See also \cite{IshKunRam03, WeiVet12}.

\vspace*{0.1cm}

Closer to our approach that entails {\it spatial} sampling, 
in a setting of distributed acoustic 
sensing and reconstruction, centralized as well as distributed 
coding schemes and sampling lattices are studied  in 
\cite{KonTelVet12}.
The rate distortion function has 
been characterized when multiple Gaussian signals from a random 
field are sampled and quantized (centralized or distributed) in 
\cite{PraNeu07}, \cite{NeuPra11,NeuPra12}.
In \cite{KashLasXiaLiu05}, a 
Gaussian field on the interval $[0,1]$ and i.i.d. 
in time, is reconstructed from compressed 
versions of $k$-sampled sequences under a mean-squared error 
criterion. The rate distortion function is studied
for schemes that reconstruct only the sampled sources
first and then reconstruct the unsampled sources by forming 
minimum mean-squared error (MMSE) estimates
based on the reconstructions for the sampled sources.
All the sampling problems  above assume a knowledge of
the underlying distribution.

\vspace*{0.1cm}

In the realm of rate distortion theory where a complete knowledge of
the signal statistics is unknown, there is a rich literature that considers
various formulations of universal coding; only a sampling is listed here. 
Directions include 
classical Bayesian and nonBayesian methods \cite{Ziv72, NeuGrDav75, NeuShi78, Rissanen84};
``individual sequences'' studies \cite{Ziv80, WeissMer01, WeissMer02a};
redundancy in quantization rate or distortion \cite{LinLugZeg95, Linder00, Linder02};
and lossy compression of noisy or remote signals \cite{LinLugZeg97, Weiss01, DemboWeiss03}. 
These works propose a variety of distortion measures to investigate
universal reconstruction performance.

\vspace*{0.1cm}

Our work differs materially from the approaches above. Sampling is
spatial rather than temporal. 
Our notion of universality involves a lack of specific knowledge
of the underlying pdf in a given compact family of pdfs. Accordingly,
in Bayesian and nonBayesian settings, we consider average
and peak distortion criteria, respectively, with an emphasis
on the former.

\vspace*{0.1cm}

Our technical contributions are as follows. 
In Bayesian and nonBayesian settings, we extend the notion
of a {\it universal sampling rate distortion function} (USRDf)
\cite{BodNar17U} to  Gaussian memoryless sources, with the objective
of characterizing the tradeoffs among sampling, compression rate and 
distortion level. To this end, we consider first the setting with known
underlying pdf, and characterize its sampling rate distortion function (SRDf). This
uses as an ingredient the rate distortion function for a discrete ``remote'' source-receiver 
model with known distribution \cite{DobTsy62, Ber71, Ber78, YamIto80}. 
When the underlying pdf is known, we show that 
the overall reconstruction can be performed -- optimally -- in two steps: the 
sampled sources are reconstructed first under a modified weighted 
mean-squared error criterion and then MMSE estimates are formed
for the unsampled sources based on the reconstructions for the sampled sources.
This is akin to the structure observed in 
\cite{BodNar17} for reconstructing discrete sources from subsets of sources
under the probability of error criterion and in
\cite{WolfZiv70} for reconstructing remote Gaussian sources.
The   USRDf for  Gaussian memoryless sources with known pdf 
will serve as a key ingredient in characterizing the USRDf for 
the Gaussian field, with known distribution, previously 
studied in \cite{KashLasXiaLiu05} in a restricted setting.
Building on the ideas developed, for the SRDf (with known pdf), 
we characterize next the USRDf for  Gaussian memoryless sources
in the Bayesian and nonBayesian settings and show that it remains optimal
to reconstruct first the sampled sources and then form estimates for 
the unsampled sources based on the reconstructions of the sampled sources.

\vspace*{0.1cm}

Our model is described in Section \ref{s:prelims} and our main results
and illustrative examples are presented in Section \ref{s:Results}.
In Section \ref{s:Proofs}, we present achievability proofs first  
when the pdf is known and then, building on it, the 
achievability proof for the universal setting, with an emphasis on the
Bayesian setting. A unified converse proof is presented thereafter.

\section{Preliminaries} \label{s:prelims}

Denote $\cM = \{ 1, \ldots, m \}$ and let 
\begin{align}
X_{\cM} = \begin{pmatrix}
             X_1 \\
             X_2 \\
             \vdots \\
             X_{m}
            \end{pmatrix} 
\end{align}
be a $\mathbbm{R}^m$-valued zero-mean (jointly) Gaussian random vector with a 
positive-definite covariance matrix.  
For a nonempty set $A \subseteq \cM $ with $|A| = k$, we denote by 
$X_{A}$ the random (column) vector $ ( X_{i}, \ i \in A )^T $,
with values in $ \mathbbm{R}^{k}$.
Denote $n$ repetitions of $X_{A}$, with values in 
$ \mathbbm{R}^{nk}$, by $X_{A}^{n} = ( X_{i}^{n}, \ i \in A)^T$.  
Each $X_{i}^{n} = ( X_{i1}, \ldots, X_{in})^T, \ i \in A,$ takes values in  ${\mathbbm{R}}^{n}$. 
Let $A^{c}  = \cM \setminus A$ and let $ \mathbbm{R}^{m}$
be the reproduction alphabet for $X_{\cM}$. All logarithms and 
exponentiations are with respect to the base 2 and all norms are $\ell_2$-norms.

\vspace*{0.1cm}

Let $ \Theta = \{ {\bf \Sigma}_{\cM \tau} \}_{\tau}${$^\dag${
\renewcommand{\thefootnote}{}\footnotetext{$^\dag$ $\Theta$ is a 
collection of covariance matrices indexed by $\tau$.
By an abuse of notation, we shall use $\tau$ to refer 
to the covariance matrix ${\bf \Sigma}_{\cM \tau}$ itself.
}}}
be a set 
of $m \times m$-positive-definite matrices, and assume 
$\Theta$ to be convex and compact in the Euclidean topology
on $\mathbbm{R}^{m \times m}$. For instance, for $m = 2$, 
\begin{align}
\qquad \Theta  = \left \{ 
\begin{pmatrix}
  \sigma_{1  }^2 &  r \sigma_{1} \sigma_{2} \\  
  r \sigma_{1 } \sigma_{2} &  \sigma_{2}^2
\end{pmatrix}
, \quad  c_1 \leq \sigma_1^2, \sigma_2^2 \leq c_2, \ -d_1 \leq r \leq d_1  \right \},
\end{align}
with $0 < c_1 \leq c_2  $ and $ 0 \leq d_1 < 1.$
Hereafter, all covariance matrices under consideration
will be taken as being positive-definite without explicit mention.
We assume $\theta$ to be a $\Theta$-valued rv with a pdf $\nu_{\theta}$
that is absolutely continuous with respect to the Lebesgue measure on
$\mathbbm{R}^{m^2}$. 
We assume $$\nu_{\theta}(\tau) > 0, \ \ \ \tau \in \Theta,$$
and that $\nu_{\theta}$ is continuous on $\Theta$.
We consider a jointly Gaussian memoryless multiple source (GMMS) $\{ X_{\cM t} 
\}_{t=1}^{\infty}$ consisting of i.i.d. repetitions of the rv 
$X_{\cM}$ with pdf known only to the extent of belonging to the 
family of pdfs $ \cP = \big \{ \nu_{X_{\cM}|\theta = \tau} =  {\cal N}({\bf 0}, 
{\bf \Sigma}_{\cM \tau})${$^\ddag${
\renewcommand{\thefootnote}{}\footnotetext{$^\ddag$Throughout this paper,
${\cal N}({\bf 0}, {\bf \Sigma})$ is used to denote the pdf 
of a Gaussian random vector with mean ${\bf 0}$ and covariance matrix ${\bf \Sigma}$.
}}},$ \ \tau \in \Theta \big  \}$.
Two settings are studied: in a Bayesian formulation, the 
pdf $\nu_{\theta}$ is taken to be {\it known}, while in a nonBayesian 
formulation $\theta$ is an {\it unknown constant} in $\Theta$.

\vspace*{0.1cm}

\vspace*{0.1cm}

\begin{definition} \label{d:RS}
For a fixed $ A \subseteq \cM $ with $|A| = k$, 
a {\it k-fixed-set} sampler ($k$-FS), 
 $1 \leq k \leq m$, collects at each $t \geq 1$,   
 $X_{A {t}}$ from $X_{\cM t}$. The output of the $k$-FS
 is $\{ X_{A t} \}_{t=1}^{\infty}$.
 
\end{definition}
 
%%%%%%%%%%%%%%%%%%%%%%%%%%%%%%%%%

\begin{definition} \label{d:encoder}
For $n \geq 1$,
an $n$-length block code with $k$-FS for a GMMS $\{ X_{\cM t}
 \}_{t=1}^{ \infty} $ with reproduction 
 alphabet $\mathbbm{R}^m$ is the pair $(  f_{n}, 
 \varphi_{n} )$ where  the encoder $f_{n}$ maps the $k$-FS output 
 $X_{A}^{n}$ into some finite set $\cJ = \{1, \ldots , J \}$ 
 and the decoder $\varphi_{n}$ maps $\cJ$ into $\mathbbm{R}^{nm}$. We 
 shall use the compact notation $(  f, \varphi),$ 
 suppressing  $n$. The rate of the code $( f, \varphi)$ with $k$-FS 
   is $\dfrac{1}{n} \log J$.

\end{definition}

%%%%%%%%%%%%%%%%%%%%%%%%%%%%%%%%%

\vspace*{0.1cm}

Our objective is to reconstruct all the components of a GMMS 
from the compressed representations of the sampled GMMS 
components under a suitable distortion criterion
with (single-letter) mean-squared error (MSE) distortion measure 
\begin{align} \label{eq:mean-squared-error}
||  x_{\cM} - y_{\cM}||^2 = \sum \limits_{i=1}^m (x_i - y_i)^2, 
 \qquad x_{\cM}, y_{\cM} \in \mathbbm{R}^m.
\end{align}
For threshold $ \Delta \geq 0, $
an $n$-length block code $(f, \varphi)$ with $k$-FS  will be required to 
satisfy one of the following $(|| \cdot ||^2, \Delta)$  distortion criterion  depending on the setting.

\vspace*{0.1cm}

\noindent (i) {\it Bayesian}: The {\it expected} distortion criterion is
\begin{align}
\begin{split}
 \label{eq:expected-fidelity-criterion-GMMS}
  \mathbb{E} \bigg [ \Big | \Big | X_{\cM}^{n} - \varphi 
 \big ( f (  X_{A}^{n} ) \big ) \Big | \Big |^2 \bigg ]  
 & =  \mathbb{E} \bigg[ \dfrac{1}{n} \sum_{t=1}^{n} 
 \Big | \Big | X_{\cM t} - \Big ( \varphi \big ( 
 f (  X_{A}^{n}) \big ) \Big )_{t}  \Big | \Big |^2 \bigg ] \\
 & = \mathbb{E} \bigg[ \mathbbm{E} \Big [ \dfrac{1}{n} \sum_{t=1}^{n}  
   \Big | \Big | X_{\cM t} - \Big ( \varphi \big ( 
 f (  X_{A}^{n}) \big ) \Big )_{  t}  \Big |\Big|^2 \Big | \theta \Big] \bigg ] \\
 & \leq  \Delta .
 \end{split}
\end{align}

\vspace*{0.1cm}
\noindent (ii) {\it NonBayesian}: The {\it peak} distortion criterion is
\begin{equation}
 \label{eq:peak-fidelity-criterion-GMMS}
  \underset{\tau \in \Theta} \sup \ \mathbb{E} 
  \bigg [  \Big | \Big | X_{\cM}^{n} -  \varphi 
 \big ( f (  X_{A}^{n} ) \big ) \Big | \Big |^2 \Big | \theta = \tau \bigg ]  
 \leq \Delta,
\end{equation} 
where $\mathbbm{E}[ \cdot | \theta = \tau]$ 
denotes $\mathbbm{E}_{ \nu_{ _{X_{\cM}^n | \theta =  \tau}} } [ \cdot ]$.

\begin{definition} \label{d:RDF}
 
A number $ R \geq 0$ is an achievable universal $k$-sample coding rate at 
  distortion level $\Delta$ if for every $\epsilon > 0$ 
and sufficiently large $n$, there exist $n$-length block codes 
with $k$-FS of rate less than $R + \epsilon$ and satisfying the 
 fidelity criterion $( || \cdot ||^2, \Delta + \ep)$ in 
\eqref{eq:expected-fidelity-criterion-GMMS}
or \eqref{eq:peak-fidelity-criterion-GMMS} above; 
$(R, \Delta)$ will be termed an achievable universal $k$-sample 
rate distortion pair under the expected or peak distortion criterion. 
The infimum of such achievable rates foe each fixed $\Delta$ 
is denoted by $R_{A}( \Delta)$.
We shall refer to $R_{A} (\Delta)$ as the {\it universal sampling 
rate distortion function} (USRDf), suppressing the dependence on 
$k$. For $|\Theta| = 1,$ the USRDf is termed simply the {\it sampling
rate distortion function} (SRDf), denoted by $\rho_{A}(\Delta).$

\end{definition}
 
\vspace*{0.1cm}

\noindent {\it Remarks}: (i) The USRDf under \eqref{eq:expected-fidelity-criterion-GMMS}
is no larger than that under \eqref{eq:peak-fidelity-criterion-GMMS}.

\vspace*{0.1cm}

\noindent (ii) When $|\Theta| = 1$, the pdf of the GMMS is, in effect, known.
  
  \vspace*{0.1cm}

\noindent Below, we recall (Chapter 1, \cite{Ihara}) the 
definition of mutual information between two random variables.
\begin{definition}
For real-valued rvs $X$ and $Y$ with a joint 
probability distribution $\mu_{X Y}$, the mutual information between 
the rvs $X$ and $Y$ is given by
\begin{align}
 I(X \wedge Y) =
 \begin{cases}
& \mathbbm{E}_{\mu_{X Y}}\Big[ \log \frac{d\mu_{X Y}}{ d\mu_X \times d\mu_Y }(X,Y)  \Big], 
\hspace*{1cm} \text{ if } \mu_{X Y} < \hspace*{-0.15cm} <  \mu_{X} \times \mu_{Y} \\
& \infty , \hspace*{5cm} \  \text{otherwise},
\end{cases}
\end{align}
where  $\mu_{X Y} < \hspace*{-0.15cm} <  \mu_{X} \times \mu_{Y}$ denotes that 
$\mu_{X Y}$ is absolutely continuous with respect to $\mu_{X} \times \mu_{Y}$
and $\frac{d\mu_{X Y}}{ d\mu_X \times d\mu_Y } $ is the Radon-Nikodym
 derivative of $\mu_{X Y}$ with respect to $\mu_{X} \times \mu_Y$.
\end{definition}

\section{Results} \label{s:Results}

We begin with a setting where the pdf of $X_{\cM}$ is known 
and provide a (single-letter) characterization for the SRDf.
Next, in a brief detour, we introduce an extension of GMMS, namely 
a  Gaussian memoryless field (GMF) and show how 
the ideas developed for a GMMS can be used to
characterize the SRDf for a GMF.
Finally, building on the SRDf for a GMMS,
a (single-letter) characterization of the USRDf  is provided 
for a GMMS in the Bayesian and nonBayesian settings.

\vspace*{0.1cm}

Throughout this paper, a recurring structural property of our achievability
proofs is this: it is optimal to reconstruct the {\it sampled }
GMMS components first under a {\it (modified) 
weighted} MSE criterion with {\it reduced threshold }
and then form deterministic (MMSE) estimates of the unsampled 
components based on the reconstruction of the former.

\vspace*{0.1cm}

Before we present our first result, we recall that for a GMMS $\{ X_{\cM t} \}_{t=1}^{\infty}$
with pdf ${\cal N}({\bf 0}, {\bf \Sigma}_{\cM})$
reconstructed under the MSE distortion criterion, the {\it standard} rate distortion function (RDf) is 
\begin{align}
 R(\Delta) & = \underset{ \substack{ \mu_{X_{\cM} Y_{\cM}} < \hspace*{-0.1cm} < \mu_{X_{\cM}} \times 
 \mu_{Y_{\cM}} \\  \mathbbm{E}[||X_{\cM} - Y_{\cM} ||^2] \leq \Delta } } \min I(X_{\cM} \wedge Y_{\cM}),
 \qquad 0 < \Delta \leq \sum \limits_{i=1}^{m} \mathbbm{E}[X_i^2] \label{eq:GMMS_RD}   \\
 & = \frac{1}{2} \sum \limits_{i=1}^{m} \Big{( \log \frac{{\lambda}_i}{\alpha} \Big)}^{+} , 
 \hspace*{3.4cm} \  0 < \Delta \leq \sum \limits_{i=1}^{m} \mathbbm{E}[X_i^2],
\end{align}
where $\lambda_{i}$s are the eigenvalues of ${\bf \Sigma}_{\cM}$,
and $\alpha$ is chosen to satisfy $ \sum \limits_{i=1}^{m} \min(\alpha, \lambda_i) = \Delta.$

\subsection{$|\Theta | = 1$: Known pdf} \label{ss:known_pdf}

Starting with $|\Theta| = 1$, 
for a GMMS $\{ X_{\cM t} \}_{t=1}^{\infty}$ with (known) pdf 
${\cal N}({\bf 0}, {\bf \Sigma}_{\cM})$, 
our first result shows that the fixed-set SRDf $\rho_{A}(\Delta)$ for a GMMS is, in effect, the 
RDf of a GMMS $\{ X_{A t} \}_{t=1}^{\infty}$
with a weighted MSE distortion measure $d_{A}$ and a reduced 
threshold; here 
$d_{A}: \mathbbm{R}^k \times \mathbbm{R}^k \rightarrow \mathbbm{R}^{+} \cup \{ 0\}$ is given by 
\begin{align} \label{eq:d_A_defn}
 d_{A}(x_{A}, y_{A}) \triangleq (x_{A} - y_{A})^T {\bf G}_{A} (x_A - y_A), 
 \qquad x_A, \ y_A \in \mathbbm{R}^k
\end{align}
with 
\begin{align} \label{eq:G_matrix}
\qquad   {\bf G}_{A} = {\bf I} + {\bf \Sigma}_{A}^{-1} 
 {{\bf \Sigma}_{{A}{{A}^c}}}  {\bf \Sigma}_{{A}{{A}^c}}^{T} {\bf \Sigma}_{A}^{-1},   
%  \quad  .
\end{align}
where ${{\bf \Sigma}_{{A}{{A}^c}}} = \mathbbm{E} [ X_{{A}} X_{A^{c}}^{T}]$.

\begin{theorem} \label{th:GMMS_SRDf}
For a GMMS $\{ X_{\cM t} \}_{t=1}^{\infty}$ with pdf 
${\cal N}({\bf 0}, {\bf \Sigma}_{\cM})$ 
and fixed $A \subseteq \cM$, the  SRDf is
\begin{align} 
\rho_{A} (\Delta) 
& = \displaystyle \min_{ \substack{ 
\mu_{X_A Y_A} < \hspace*{-.1cm} < \mu_{X_A} \times \mu_{Y_{A}} \\
\mathbb{E} [ d_{A}(X_{A}, Y_{A}) ] \leq {\Delta} - {\Delta}_{ \min, A}}} 
I \big( X_{A} \wedge Y_{A} \big), \hspace*{3.1cm} \ \ 
\Delta_{ \min,A} < \Delta \leq \Delta_{\max} \label{eq:SRDf_GMMS} \\
&  =  \displaystyle \frac{1}{2} \sum_{i=1}^{k}
\Big{( \log \frac{{\lambda}_i}{\alpha} \Big)}^{+},  \hspace*{6.4cm}
\Delta_{ \min,A} < \Delta \leq \Delta_{\max} \label{eq:R_D_GMMS_eigval}
\end{align}
where
\begin{align}
{\Delta}_{ \min, A} =   \sum_{i \in {A}^{c}} \big( 
\mathbbm{E} [X_i^2] - \mathbbm{E} [X_i X_{A}^T ] {\bf \Sigma}_{A}^{-1} 
\mathbbm{E} [ X_{A} X_i] \big), \quad {\Delta}_{\rm max} =  \sum_{i \in \cM} 
\mathbbm{E} [X_i^2 ]
\end{align}
and ${\lambda}_i$s are the eigenvalues of ${\bf G}_{A} {\bf \Sigma}_{A}$,
and $ \alpha $ is chosen to satisfy $ \displaystyle \sum_{i=1}^{k} \min 
({\alpha}, {\lambda}_i) = {\Delta} - {\Delta}_{\min,A}$. 

\end{theorem}

Comparing \eqref{eq:SRDf_GMMS} with \eqref{eq:GMMS_RD}, it can be
seen that \eqref{eq:SRDf_GMMS} is, in effect, the
RDf for a GMMS with weighted MSE distortion measure.
In contrast to the RDf \eqref{eq:GMMS_RD}, in \eqref{eq:SRDf_GMMS} the minimization
involves only $X_A$ (and not $X_{\cM}$)
under a weighted MSE criterion with reduced threshold level. 
For $k = m$, i.e., $A = \cM$, however this reduces to the 
RDf \eqref{eq:GMMS_RD}. 
Also, for every feasible distortion level
the SRDf for any $A \subset \cM$ is no smaller than that with $A = \cM$.

\vspace*{0.2cm}

In Section \ref{s:Proofs}, the achievability proof of the theorem above
involves reconstructing the sampled components of the GMMS first, and then 
forming MMSE estimates for the unsampled components based on the former.
Accordingly, in \eqref{eq:SRDf_GMMS}, the MSE in the reconstruction of the entire GMMS 
is captured jointly  by the weighted MSE ({with weight-matrix} ${\bf G}_{A}$) 
in the reconstructions of the sampled components 
and the minimum distortion $\Delta_{\min,A}$.

\vspace*{0.2cm}

Observing that \eqref{eq:SRDf_GMMS} is equivalent to the RDf of a GMMS
with a weighted MSE distortion measure enables us to provide an analytic expression
for the SRDf using the standard reverse water-filling 
solution \eqref{eq:R_D_GMMS_eigval} \cite{Ihara}. An instance of this is shown in
the example below.

\begin{example} \label{ex:GMMS_SRDf}
For a GMMS with a $k$-FS with $k = 1$, this example illustrates 
the effect of the choice of the sampling set on SRDf.
Consider a GMMS $\{X_{\cM t} \}_{t=1}^{\infty}$ with 
covariance matrix ${\bf \Sigma}_{\cM}$ given by
\begin{align}
\quad 
{\bf \Sigma}_{\cM} = \begin{pmatrix}
  \sigma_1^2 & r_{12} \sigma_1 \sigma_2  & \cdots & r_{1m} \sigma_1 \sigma_m \\
  r_{21} \sigma_1 \sigma_2 & \sigma_2^2 & \cdots & r_{2m} \sigma_2 \sigma_m \\
  \vdots  & \vdots  & \ddots & \vdots  \\
  r_{m1} \sigma_1 \sigma_m & r_{m2} \sigma_2 \sigma_m & \cdots & \sigma_m^2
\end{pmatrix},
\end{align}
where $r_{ij} = r_{ji}, \ 1 \leq i , j \leq m.$
For $A = \{j\}, \ j = 1, \ldots,m$, we have 
\begin{align}
 {\bf G}_{ \{j\} } {\bf \Sigma}_{ \{j\}}
 =   \big( 1 + \sum \limits_{i \neq j} \frac{r_{ij}^2 \sigma_{i }^2}{\sigma_j^2} \big ) 
   \sigma_{ {j }}^2 
 =   \sigma_{ {j }}^2  +  \sum \limits_{i \neq j} r_{ij}^2 \sigma_{i }^2 
\end{align}
and hence from \eqref{eq:R_D_GMMS_eigval}, the SRDf is
\begin{align} \label{eq:ex1_SRDf_k1}
 \rho_{ \{j\} }(\Delta) 
 & = \frac{1}{2} \log \left ( 
 \frac{ \sigma_{ {j }}^2  +  \sum \limits_{i \neq j} r_{ij}^2 \sigma_{i }^2  }{ \Delta - \Delta_{\min, \{j\}} } \right ) \\
 & = \frac{1}{2} \log \left ( \frac{ \sum \limits_{i=1}^{m} 
      \sigma_i^2 - \Delta_{\min, \{j\} }  }{ \Delta - \Delta_{\min, \{j\} }} \right )
\end{align}
for $\Delta_{\min, \{j\} } < \Delta \leq \sum \limits_{i=1}^{m} \sigma_i^2 ,$
where $\Delta_{\min, \{j\} } = \sum \limits_{i \neq j}^{m} \sigma_i^2 (1- r_{ij}^2)$. 
Observe that every SRDf $ \rho_{ \{j \} }(\Delta)$ is a monotonically increasing 
function of $\Delta_{\min, \{j\} }$
and that the SRDfs are translations of each other and hence decrease at the same rate.
Thus, the SRDf with the smallest $\Delta_{\min, \{j\}}$ 
is uniformly best among all fixed-set SRDfs.
For $k > 2$ however, there may not be any $A \subset \cM, \ |A|=k,$ 
whose fixed-set SRDf is uniformly best for all distortion levels.
\qed

\end{example}

%%%%%%%%%%%%%%%%%%%%%%

Before turning to the USRDf for a GMMS, the ideas involved in
Theorem \ref{th:GMMS_SRDf} are used to study sampling and lossy compression 
of a Gaussian field which affords greater flexibility in the choice of sampling set.
While Gaussian fields have been studied extensively under different formulations,
we consider a
Gaussian memoryless field (GMF) as in \cite{KashLasXiaLiu05}, which is described next.
In lieu of $\cM$ and Gaussian rv $X_{\cM}$ in Section \ref{s:prelims}, consider 
$I = [0,1] \subset \mathbbm{R}$ and let $X_{I} = \{X_u, \ u \in I \}$ be a
$\mathbbm{R}^I \triangleq \{ \mathbbm{R}, \ u \in I \} $-valued zero-mean Gaussian 
process$^\dag${\renewcommand{\thefootnote}{}\footnotetext{$^\dag$A 
Gaussian process on an interval $[0,1]$ means that any finite collection of rvs 
$(X_{s_{1}}, \ldots, X_{s_{l}}), \ s_{i} \in [0,1], \ i \in \{ 1, \ldots, l \}, 
\ l \in \mathbbm{N},$ are jointly Gaussian.}} with a bounded covariance function 
$r(s_1,s_2) = \mathbbm{E}[ X_{s_{1}} X_{s_{2}}] , \  s_1, s_2 \in I$, 
such that, for any finite  $B \subset I$
\begin{align}
 \mathbbm{E}[X_{B} X_{B}^T]
\end{align}
is a positive-definite matrix and
\begin{align} \label{eq:GMF_r_integ}
 \int \limits_{I} \int \limits_{I} |r(u,v)| \, du \,dv < \infty.
\end{align}
A GMF$^\ddag${\renewcommand{\thefootnote}{}\footnotetext{
$^\ddag$Extensive studies of memoryless repetitions of a Gaussian process 
exist, cf. \cite{KashLasXiaLiu05}, \cite{NeuPra11}, 
under various terminologies.}} $\{ X_{I t}  \}_{t=1}^{\infty}$ 
consists of i.i.d. repetitions of $X_{I}$. We consider a GMF sampled finitely 
by a $k$-FS at $A \subset I$, with $|A| = k$, and with a reconstruction 
alphabet $\mathbbm{R}^I$.

\vspace*{0.2cm}

\noeqref{eq:field-mean-squared-error}

For a GMF with fixed-set sampler and MSE distortion measure
\begin{align}\label{eq:field-mean-squared-error}
  \qquad  || x_{I} - y_{I} ||^2 &  = \int \limits_{ I} (x_u - y_u)^2 \, du, 
 \qquad x_{I}, y_{I} \in \mathbbm{R}^I,
\end{align}
the sampling rate distortion function is defined 
as in Definitions \ref{d:encoder} and \ref{d:RDF} with the 
decoder $\varphi$ characterized by a collection of mappings
$\varphi  = \{ \varphi_{u} \}_{u \in I}$ with
\begin{align}
 \varphi_u : \{1, \ldots,J \} \rightarrow \mathbbm{R}^{n}, \ \ u \in I.
\end{align}
Analogous to a GMMS,
for a GMF sampled at $A =  \{ a_1, \ldots, a_k \},$
$ 0 \leq a_i \leq 1, \ i = 1, \ldots,k,$
our next result shows that the SRDf is, in effect, the RDf of a
GMMS $\{ X_{A t} \}_{t=1}^{\infty}$ with a 
weighted MSE distortion measure with weight-matrix given by
\begin{align} \label{eq:GMF_G}
 {\bf G}_{A, I} =  {\bf \Sigma}_{A}^{-1} \Big ( \int\limits_{I}
   \mathbbm{E}[X_{A} X_{u}]  \mathbbm{E}[X_{u} X_{A}^T]   \,du \Big ) {\bf \Sigma}_{A}^{-1},
\end{align}
with $\int $ connoting element-wise integration. 
Note that for every $ 0 \leq s_1, s_2 \leq 1$,
\eqref{eq:GMF_r_integ} and the boundedness of $r(\cdot)$
imply that the integral
\begin{align}
 \int \limits_{I} r(u,s_1) r(u,s_2) \, du
\end{align}
exists and hence \eqref{eq:GMF_G} is well-defined.

\begin{proposition} \label{prop:SRDf_GMF}
 For a GMF $\{ X_{It} \}_{t=1}^{\infty}$  with   $A \subset I$,  the SRDf is 
\begin{align}
\label{eq:SRDf-gauss-field}
\rho_{A} (\Delta) & =  
\displaystyle \min_{ \substack{\mu_{X_{A}Y_{A}}  < \hspace*{-0.1cm} < \mu_{X_{A}} \times \mu_{Y_{A}} \\
\mathbb{E} [{(X_{A} - Y_{A})}^{T} {\bf G}_{A, I}  (X_{A} - Y_{A})]
\leq   {\Delta}  - {\Delta}_{\min, A}} }
I \big( X_{A} \wedge Y_{A} \big), \qquad
\mbox{}{\Delta}_{ \min,A} < {\Delta} \leq {\Delta}_{\rm max} \\
&  =  \frac{1}{2} \sum_{i = 1}^{k} 
\Big{( \log \frac{{\lambda}_i}{ \alpha } \Big)}^{+}, \hspace*{5.7cm}
\mbox{}{\Delta}_{ \min, A} < {\Delta} \leq {\Delta}_{\rm max}
\label{eq:r-d-gauss-field}
\end{align}
where
\begin{align} \label{eq:GMF_D_min}
{\Delta}_{{\rm min }, A} & =  \int\limits_{I}
\big ( \mathbbm{E}[X_{u}^2] -  \mathbbm{E}[X_{u} X_{A}^T ]
{\bf \Sigma}_{A}^{-1} 
\mathbbm{E}[X_{A} X_{u} ] \big )
\,du  \quad \text{ and } \quad   
{\Delta}_{\rm max} =   \int \limits_{I} \mathbbm{E}[X_{u}^2] \,du ,
\end{align}
and ${\lambda}_i$s are the eigenvalues of ${\bf G}_{A,I} {\bf \Sigma}_{A}$, 
and $ \alpha $ satisfies $\displaystyle \sum_{i  = 1}^{k}
\min ({\alpha}, {\lambda}_i) = \Delta - \Delta_{\min,A} $.

\end{proposition}

%%%%%%%%%%%%%%%%%%%%%%

\vspace*{0.15cm}

The SRDf for a GMF \eqref{eq:SRDf-gauss-field} and its equivalent form
\eqref{eq:r-d-gauss-field} can be seen as  counterparts of 
\eqref{eq:SRDf_GMMS} and \eqref{eq:R_D_GMMS_eigval},
with \eqref{eq:r-d-gauss-field} being the reverse
water-filling solution for \eqref{eq:SRDf-gauss-field}.
As before, the expression \eqref{eq:SRDf-gauss-field}
is the RDf of a GMMS with a weighted MSE 
distortion measure. In Section \ref{s:Proofs}, 
an achievability proof for the proposition above is provided by
adapting the ideas developed for Theorem \ref{th:GMMS_SRDf};
a converse proof for the proposition is provided involving a set of techniques 
different from the converse proof provided for Theorem \ref{th:GMMS_SRDf}.

\vspace*{0.1cm}

In contrast to a GMMS with a discrete set $\cM$, 
for a GMF,  $I$ being an interval affords greater flexibility
in the choice of the sampling set allowing for a better understanding 
of the structural properties of the ``best'' sampling set.
In contrast to Example \ref{ex:GMMS_SRDf} in the example below, 
considering a GMF with a stationary Gauss-Markov process, 
we show the structure of the optimal set for minimum distortion
for $k>2$ as well. In general, the optimal sampling set is a function of the
threshold $\Delta$.

\begin{example}
 Consider a GMF with a zero-mean, stationary Gauss-Markov process $X_I$ over $I = [0,1]$
 with covariance function
 \begin{align}
\quad  r(s,u)   = p^{|s-u|}, \quad 0 \leq s,u  \leq 1,
 \end{align}
and $0 < p < 1$. Note that the correlation between any two points
in the interval depends only on the distance between them.
For the Gauss-Markov process $X_I$, for any
$0 \leq u_1 < u_2 < \cdots < u_l \leq 1, \ l >2$, it holds that
\begin{align} \label{eq:ex2_markov_prop}
 X_{u_1} \MC X_{u_2} \MC \cdots \MC X_{u_l}.
\end{align}
For a $k$-FS with $k=1$ and $A = \{a\}, \ 0 \leq a \leq 1$,
\begin{align}
{\bf G}_{ \{ a \},I } = 1 - \Delta_{\min, \{a \} }
\end{align}
and $\mathbbm{E}[X_a^2] = 1$.
In \eqref{eq:r-d-gauss-field}, the eigenvalue $\lambda_1$ 
is $ {\bf G}_{\{a\}, I} {\bf \Sigma}_{ \{a\}} = 1 - \Delta_{\min, \{a \} }$ itself 
and hence, the SRDf is 
\begin{align} \label{eq:ex2_SRDf_k1}
 \rho_{\{a\}}(\Delta) 
 = \frac{1}{2} \log \frac{ 1 - \Delta_{\min, \{a\} } }{\Delta  - \Delta_{\min, \{a\} } }
\end{align}
for $\Delta_{\min, \{a\}}  < \Delta \leq 1$, where 
\begin{align}
\Delta_{\min, \{ a\} } 
& = \int_{ 0 }^a \Big( \mathbbm{E}[X_{u}^2] - \frac{\mathbbm{E}^2[X_{u}X_{a}]}{\mathbbm{E}[X_a^2]} \Big ) \, du 
+  \int_{ a }^1 \Big( \mathbbm{E}[X_{u}^2] - \frac{\mathbbm{E}^2[X_{u} X_{ a}]}{\mathbbm{E}[X_a^2]} \Big ) \, du \\
& = \int_{ 0 }^a \big ( 1 -  p^{2(a-u)} \big ) \, du  +  \int_{ a }^1 \big( 1 - p^{2(u-a)} \big ) \, du \\
& = 1 - \frac{p^{2 a} - 1 + p^{2(1-a)} - 1}{ \ln {p}}.
\end{align}
Note that the SRDf $\rho_{ \{a\} }(\Delta)$ is a monotonically increasing function of 
$\Delta_{\min, \{a\} }$, which in turn is a monotonically increasing function of $| a - 0.5 |$.
Thus, 
$ \rho_{ \{0.5\} }(\Delta)$ is uniformly best among all 
SRDfs $ \rho_{ \{a\}}(\Delta)$, $0 \leq a \leq 1$, for all distortion levels.
Now, for a $k$-FS with $k>2$ and $A = \{  a_1 = 0, a_2, \ldots, a_{k-1}, a_{k} = 1 \}$,
with $a_{i} \leq a_{i+1}, \ i = 1, \ldots,k-1,$
the minimum distortion  $\Delta_{\min,A} $ admits a simple form
\begin{align}
 \Delta_{\min, A} = 1 - \sum \limits_{i=1}^{k-1} \gamma(a_{i+1} - a_{i}),
\end{align}
where $\gamma(a_{i+1} - a_{i})$ is according to
\begin{align}
\quad \gamma(a) \triangleq \frac{1}{1 - p^{2 a}} \Big ( \frac{ p^{2 a}(1- 2a \log p) - 1 }{\log p}  \Big),
\quad 0 < a < 1.
\end{align}
The minimum reconstruction error $\Delta_{\min, A} $ is the ``sum'' of the 
 minimum error in reconstructing each segment $[a_i, a_{i+1}]$ of the 
 GMF. Now, the Markov property \eqref{eq:ex2_markov_prop} implies that the 
minimum error in reproducing each component $u \in I$ 
is determined by its nearest sampled points and hence
the minimum error in reconstructing each segment $[a_i, a_{i+1}]$
of the GMF is independent of the location of sampling points other than $a_i$, $a_{i+1}$ 
and is given by
$$(a_{i+1} - a_{i}) - \gamma(a_{i+1} - a_i).$$ 
The stationarity of the field means that this
minimum error depends on the length $|a_{i+1} - a_{i}|$ alone.
Observing that $\gamma(a)$ is a concave function of $a$ over $(0,1]$,  $\Delta_{\min,A}$ above
is seen to be minimized when $ a_{i+1} - a_{i} = \frac{1}{k-1}, \ i = 1, \ldots,k-1$, i.e., 
when the sampling points are spaced uniformly. However, such a placement is not
optimal for all distortion levels.
\qed

\end{example}

\vspace*{0.1cm}

%%%%%%%%%%%%%%%%%%%%%%%%%%%%%%%%%%%%%%%%%%%%

%%%%%%%%%%%%%%%%%%%%%%%%%%%%%%%%%%%%%%%%%%%%

\subsection{Universal setting} \label{ss:universal}

Turning to the universal setting with a GMMS, 
consider a set $\Theta_1 = \{{\bf \Sigma}_{A \tau}, \ \tau \in \Theta\} 
\subset \mathbbm{R}^{k^2}$ with $\tau_1 \in \Theta_1$ indexing the members 
of $\Theta_1$, i.e., $\Theta_1 = \{{\bf \Sigma}_{A \tau_1}  \}_{\tau_1}$
{$^\dag${
\renewcommand{\thefootnote}{}\footnotetext{$^\dag$ The collection of covariance
matrices ${\bf \Sigma}_{A \tau}$ are indexed by $\tau_1$ and by an abuse 
$\tau_1$ will also be used to refer to ${\bf \Sigma}_{A \tau_1}$ itself.
}}}. 
An encoder $f$ associated with a $k$-FS observes $X_{A}^n$ alone and 
cannot distinguish among jointly Gaussian pdfs in $ \cP$ that have the 
same marginal pdf $ \nu_{X_{A}|\theta = \tau}$. 
Accordingly (and akin to \cite{BodNar17U}),
consider a partition of $\Theta$ comprising ``ambiguity'' atoms,
with each atom of the partition comprising  $\tau$s with identical 
$ \nu_{X_{A}|\theta = \tau}$, i.e., identical ${\bf \Sigma}_{A \tau}$
and for each $\tau_1 \in \Theta_1$, $\Lambda(\tau_1)$ is the collection 
of $\tau$s in the ambiguity atom indexed by $\tau_1$, i.e.,
 \begin{align}
 \qquad {\bf \Sigma}_{A \tau_1} \triangleq {\bf \Sigma}_{A  \tau}, \quad \tau \in \Lambda(\tau_1).
 \end{align}
Let $\theta_1$ be a $\Theta_1$-valued rv
induced by $\theta$.
It is easy to see that $\Theta_1$ and $\Lambda(\tau_1), \ \tau_1 \in \Theta_1,$ are 
convex, compact subsets of $\mathbbm{R}^{k^2}$ and 
the rv $\theta_1$ admits a pdf $\nu_{\theta_1}$ induced by $\nu_{\theta}$.

\vspace*{0.2cm}

\noindent In the Bayesian setting,
\begin{align}
 \nu_{X_{A}|\theta_1 = \tau_1} = \nu_{X_{A}|\theta = \tau}
 = {\cal N}({\bf 0}, {\bf \Sigma}_{A  \tau_1}), \ \quad \tau \in \Lambda(\tau_1).
\end{align}
In the nonBayesian setting, in order to retain the same notation, we choose
$\nu_{X_{A}|\theta_1 = \tau_1}$ to be the right-side above.

\vspace*{0.1cm}

Our characterization of the USRDf builds
on the structure of the SRDf for a GMMS.
Accordingly, in the Bayesian setting,
consider the set of (constrained) probability measures
\begin{align}
 \kappa_{A}^{\cB}(\delta, \tau_1) & \triangleq \{ \mu_{\theta X_{\cM} Y_{\cM}}:
 \ \theta, X_{\cM} \MC \theta_1, X_{A}  \MC Y_{\cM}, \
 \ \mu_{X_{A} Y_{\cM} | \theta_1 = \tau_1} 
 < \hspace*{-0.15cm} < \mu_{X_{A} | \theta_1 = \tau_1} \times \mu_{Y_{\cM} | \theta_1 = \tau_1},  \\
 & \hspace*{3.5cm} \mathbbm{E}[ || X_{\cM} - Y_{\cM}||^2 |\theta_1 = \tau_1 ] \leq 
 \delta \}
\end{align}
and (constraint) minimized mutual information
\begin{align} \label{eq:prim_Bayesian}
 \rho_{A}^{\cB}(\delta, \tau_1) \triangleq  \underset{\kappa_{A}^{\cB}(\delta, \tau_1) } \min \ 
 I(X_A \wedge Y_{\cM} | \theta_1 = \tau_1).
\end{align}
Correspondingly, in the nonBayesian setting, consider 
\begin{align}
 \kappa_{A}^{n\cB}(\delta, \tau_1) & \triangleq \{ \mu_{X_{\cM} Y_{\cM} | \theta = \tau}:
 \  \mu_{Y_{\cM}|X_{\cM}, \theta = \tau} \! = \! \mu_{Y_{\cM}|X_{A}, \theta_1 = \tau_1}, \ 
%  \tau \in \Lambda(\tau_1), \
 \mu_{X_{A} Y_{\cM} | \theta_1 = \tau_1} \! < \hspace*{-0.15cm} < \! \mu_{X_{A} | \theta_1 = \tau_1} \times 
 \mu_{Y_{\cM} | \theta_1 = \tau_1}
 %P_{Y_{A^c}| Y_A, \theta_1 = \tau_1}
 , \\ 
& \hspace*{3cm} \mathbbm{E}[ ||X_{\cM} - Y_{\cM}||^2 |\theta = \tau ] \leq  \delta, \ \tau \in \Lambda(\tau_1) \}
\end{align}
and 
\begin{align} \label{eq:prim_nonBayesian}
 \rho_{A}^{n \cB}(\delta, \tau_1) \triangleq  \underset{\kappa_{A}^{n \cB}(\delta, \tau_1) } \inf \ 
 I(X_A \wedge Y_{\cM} | \theta_1 = \tau_1).
\end{align}

\vspace*{0.1cm}

\noindent {\it Remark}: In \eqref{eq:prim_Bayesian} and \eqref{eq:prim_nonBayesian},
the minimization is with respect to the conditional measure $\mu_{Y_{\cM}|X_{A},\theta_1 = \tau_1}$.

\vspace*{0.15cm}

The minimized conditional mutual informations above will
be a key ingredient in the characterization of USRDf.
First, we show in the proposition below that \eqref{eq:prim_Bayesian} 
and \eqref{eq:prim_nonBayesian} admit simpler forms involving
rvs corresponding to the sampled components of the GMMS and 
their reconstruction alone. In the Bayesian setting, 
for each $\tau_1 \in \Theta_1$,
the mentioned simpler form involves a weighted MSE
distortion measure $d_{A \tau_1}$
with weight-matrix ${\bf G}_{A, \tau_1}$, defined 
as in \eqref{eq:G_matrix}  with $ {{\bf \Sigma}_{{A}{{A}^c}}}$
replaced by $\mathbbm{E}[X_{A} X_{A^c}^T | \theta_1 = \tau_1]$
and
\begin{align} \label{eq:d_A_tau_1}
 d_{A \tau_1}(x_{A}, y_{A}) \triangleq (x_{A} - y_{A})^T {\bf G}_{A, \tau_1} 
 (x_A - y_A), \qquad x_A, y_A \in \mathbbm{R}^k.
\end{align}
In the Bayesian setting, the modified distortion measure $d_{A \tau_1}$ 
plays a role similar to that of $d_A$. 

\vspace*{0.2cm}

\noindent {\it Remark}: 
Clearly, $\rho_A^{n \cB}(\delta, \tau_1)$ is a nonincreasing function of $\delta >
\Delta_{\min,A,\tau_1}.$ Convexity of $\rho_A^{n \cB}(\delta, \tau_1)$ can be shown
as in \cite{Sakrison}, and convexity implies the continuity of $\rho_A^{n \cB}(\delta, \tau_1)$.
Now, to show the convexity, pick any $\delta_1, \ \delta_2 > \Delta_{\min,A, \tau_1}
 $ and $ \ep > 0. $ For $i = 1, 2,$ let $\mu^{i} \in \kappa_{A}^{n \cB}(\delta_i, \tau_1) 
 $ be such that 
 \begin{align}
 I_{\mu^{i} } ( X_{A} \wedge Y_{\cM} | \theta_1 = \tau_1 ) 
 \leq \rho_{A}^{n \cB}(\delta_i) + \ep.
\end{align}
For $\alpha > 0,$
by the standard convexity arguments, it can be seen that 
$\alpha \mu^1 + (1-\alpha)\mu^2 \in \kappa_{A}^{n \cB}(\alpha \delta_1 + (1-\alpha) \delta_2, \tau_1)$
and  
\begin{align} \label{eq:convexity_nonB_prim}
 I_{ \alpha  \mu^{1} + (1-\alpha) \mu^{2} } ( X_{A} \wedge Y_{\cM} | \theta_1 = \tau_1 ) 
 \leq \alpha \rho_{A}^{n \cB}(\delta_1) + (1-\alpha) \rho_{A}^{n \cB}(\delta_2) +   \ep.
\end{align}
Since \eqref{eq:convexity_nonB_prim} holds for any $\ep > 0$, in the limit, we have
\begin{align}
 \rho_{A}^{n \cB}(\alpha \delta_1 + (1-\alpha)  \delta_2 )
 \leq \alpha \rho_{A}^{n \cB}(\delta_1) + (1-\alpha) \rho_{A}^{n \cB}(\delta_2).
\end{align}

\vspace*{0.1cm}

\begin{proposition} \label{prop:prim_USRDf_equiv_form}
For each $\tau_1 \in \Theta_1$,  in the Bayesian setting 
 \begin{align}
 \rho_A^{\cB}(\delta, \tau_1)
 = \underset{ \mu_{X_{A} Y_{A} | \theta_1 = \tau_1} < \hspace*{-0.1cm} < 
 \mu_{X_{A} | \theta_1 = \tau_1} \times \mu_{ Y_{A} | \theta_1 = \tau_1} \atop
 \mathbb{E} [ d_{A \tau_1}(X_{A}, Y_{A}) |\theta_1 = \tau_1 ] 
  \leq {\delta} - {\Delta}_{ \min, A, \tau_1} }
 \min  I \big( X_{A} \wedge Y_{A} | \theta_1 = \tau_1 \big) 
 \label{eq:Bay_prim_equiv_form}
 \end{align}
for $ \delta > \Delta_{ \min,A,\tau_1} $, where
\begin{align}
 \Delta_{\min,A, \tau_1} =  
 \mathbbm{E} \big [ \mathbbm{E} \big[ \underset{y_{A^c} \in \mathbbm{R}^{m-k} } 
 \min \sum \limits_{i \in A^c} (X_i - y_i)^2 
 |X_{A}, \theta_1 = \tau_1 \big] \big| \theta_1 = \tau_1 \big].
\end{align}
For each $\tau_1 \in \Theta_1$,  in the nonBayesian setting 
\begin{align}
\qquad  \rho_A^{n \cB}(\delta, \tau_1) = \underset{ 
%     \atop 
   \mathbbm{E}[ || X_{\cM} - Y_{\cM}||^2 |\theta = \tau ] \leq  \delta, \ \tau \in \Lambda(\tau_1)  } 
 \inf  I \big( X_{A} \wedge Y_{A} | \theta_1 = \tau_1 \big) , \quad \delta >  \Delta_{ \min,A,\tau_1}  ,
 \label{eq:nonBay_prim_equiv_form}
 \end{align}
where the infimum in \eqref{eq:nonBay_prim_equiv_form} is over 
$\mu_{Y_{\cM}|X_{\cM}, \theta  = \tau},  $ such that
\begin{align}
\qquad \mu_{Y_{\cM}|X_{\cM}, \theta = \tau} & = \mu_{Y_{A}|X_{A}, \theta_1 = \tau_1} \times
 \mu_{Y_{A^c}| Y_A, \theta_1 = \tau_1}, \ \tau \in \Lambda(\tau_1), \ \text{ and } \\
 \mu_{X_{A} Y_{A}| \theta_1 = \tau_1} & < \hspace*{-0.15cm} < \mu_{X_{A}| \theta_1 = \tau_1}
 \times \mu_{Y_{A}| \theta_1 = \tau_1} 
\end{align}
and
\begin{align}
 \Delta_{\min,A, \tau_1} = 
 \underset{ \mu_{Y_{A^c}|X_A , \theta  = \tau} = \mu_{Y_{A^c}|X_A , \theta_1 = \tau_1} } \inf \
 \underset{\tau \in \Lambda(\tau_1)}
 \max \sum \limits_{i \in A^c} \mathbbm{E}[ (X_i - Y_i)^2 | \theta = \tau].
\end{align}

\end{proposition}

\vspace*{0.1cm}

\noindent {\it Remark}: From \eqref{eq:Bay_prim_equiv_form}, notice that
$\rho_A^{\cB}(\delta, \tau_1)$ is, in effect, the rate distortion function for a GMMS with 
pdf $\nu_{X_{A}|\theta_1 = \tau_1}$ and weighted MSE distortion measure. 
Hence, the minimum in \eqref{eq:Bay_prim_equiv_form} and ergo that in \eqref{eq:prim_Bayesian}
exist and the standard properties of a rate distortion function 
are applicable to $\rho_A^{\cB}(\delta, \tau_1)$ as well, i.e.,
$\rho_A^{\cB}(\delta, \tau_1)$ is a convex, nonincreasing, continuous function of 
$ \delta > \Delta_{\min, A, \tau_1}  $. % \leq \Delta_{\max}.$

\vspace*{0.2cm}

%%%%%%%%%%%%%%%%%%%%%%
 
\begin{theorem} \label{th:GMMS_USRDf}
 For a GMMS $\{X_{\cM t} \}_{t=1}^{\infty}$ with fixed $A \subseteq \cM$,
the Bayesian USRDf is
\begin{align} \label{eq:USRDf_Bayesian}
 R_{A} (\Delta) =  \underset{ \{\Delta_{\tau_1}, \ \tau_1 \in \Theta_1 \} \atop 
 \mathbbm{E}[\Delta_{\theta_1}] \leq \Delta } \min \underset{\tau_1 \in \Theta_1} \max \ 
 \rho_A^{\cB}(\Delta_{\tau_1}, \tau_1)
\end{align}
for $\Delta_{\min,A} < \Delta \leq \Delta_{\max}$, where 
\begin{align}
 \Delta_{\min,A} =  \mathbbm{E} \big[ \mathbbm{E} \big[ \underset{y_{A^c} \in \mathbbm{R}^{m-k} } 
 \min \sum \limits_{i \in A^c} (X_i - y_i)^2 |X_{A}, \theta_1 \big] \big ]
 \ \ \text{ and } \ \
 \Delta_{\max} = \sum \limits_{i \in \cM} \mathbbm{E} [ X_i^2 ].
\end{align}
The nonBayesian USRDf is
 \begin{align} \label{eq:USRDf_nonBayesian}
  R_{A} (\Delta) = \underset{\tau_1 \in \Theta_1} \max \ 
  \rho_{A}^{n \cB}(\Delta, \tau_1)
 \end{align}
 for $\Delta_{\min,A} < \Delta \leq \Delta_{\max}$, where 
 \begin{align}
  \Delta_{\min,A} = \underset{\tau_1 \in \Theta_1} \sup 
  \underset{ \mu_{Y_{A^c}|X_A , \theta = \tau} = \mu_{Y_{A^c}|X_A , \theta_1 = \tau_1} } \inf 
  \ \underset{\tau \in \Lambda(\tau_1)}
 \max \sum \limits_{i \in A^c} \mathbbm{E}[ (X_i - Y_i)^2 | \theta = \tau]
   \ \ \text{ and } \ \ 
   \Delta_{\max} =  \underset{\tau \in \Theta} \max \sum \limits_{i=1}^m 
   \mathbbm{E}[X_i^2|\theta = \tau].
 \end{align}
 
\end{theorem}

\noindent {\it Remark}:
In Appendix \ref{app:proof_min_max_Bay} a simple proof (using contradiction arguments)
is provided to show the existence of  $\{\Delta_{\tau_1} , \ \tau_1 \in \Theta_1 \}$, 
with $\Delta_{\tau_1}$ being continuous in $\tau_1$, that attains 
the minimum and the maximum in \eqref{eq:USRDf_Bayesian}.

\vspace*{0.2cm}

Notice that $\rho_{A}^{\cB}(\delta, \tau_1)$ and $\rho_{A}^{ n \cB}(\delta, \tau_1)$ are 
reminiscent of the SRDf for a GMMS and, in fact, reduce to the SRDf for 
a GMMS with $ \nu_{X_{\cM}| \theta = \tau} $ 
for $\tau \in \Lambda(\tau_1)$  when $\Lambda(\tau_1)$ is a singleton. 
Thus, the equivalent forms \eqref{eq:Bay_prim_equiv_form} and 
\eqref{eq:nonBay_prim_equiv_form} can be seen as counterparts of \eqref{eq:SRDf_GMMS}. 
Additionally, in Section \ref{s:Proofs}, we show that 
$\rho_{A}^{\cB}(\delta,\tau_1)$ and $\rho_{A}^{n\cB}(\delta,\tau_1)$
are continuous in $\tau_1 \in \Theta_1$.

\vspace*{0.2cm}

The Bayesian USRDf with an outer minimization over 
$\{ \Delta_{\tau_1}, \ \tau_1 \in \Theta_1 \}$ can be strictly smaller than
its nonBayesian counterpart. An illustration of the comparison of the 
Bayesian and nonBayesian USRDfs is provided in the example below.

\begin{example}
For $\cM = \{1,2\}$ and fixed
$\sigma^2 > 0, \ r_{\min} > 0 $ and $ r_{\max} < 1$,
consider a GMMS with
pdf in $\Theta$, where
each $ \Theta = \{ {\bf \Sigma}_{\cM \tau} \}_{\tau}$,
where each ${\bf \Sigma}_{\cM \tau}$ is given by
\begin{align}
\quad 
{\bf \Sigma}_{\cM \tau}
 =   \begin{pmatrix}
  \sigma^2 &  r_{ \tau} \sigma^2 \\  
  r_{ \tau } \sigma^2 &  \sigma^2
\end{pmatrix}
\end{align}
for $ r_{\min} \leq  r_{\tau } \leq r_{\max}, \ \tau \in \Theta$.
Let $\theta$ be a $\Theta$-valued rv with  pdf $\nu_{\theta}$
continuous on $\Theta$.
For a $k$-FS with $k = 1$, 
for both $A = \{1\}$ and $A = \{ 2\}$, $\Theta_1$ is a singleton.
Hence, in the Bayesian setting, the minimum and maximum in
\eqref{eq:USRDf_Bayesian} are vacuous.
For $A = \{1\}, \ \{2\},$ in the Bayesian setting we have 
\begin{align}
 & {\bf G}_{A, \tau_1} = 1 + \mathbbm{E}^2[r_{\theta}] , \\
 & \Delta_{\min,A, \tau_1} = \sigma^2(1 - \mathbbm{E}^2 [r_{\theta}  ]),
\end{align}
and \eqref{eq:USRDf_Bayesian} now 
yields the Bayesian USRDf to be
\begin{align} \label{eq:ex3_bayesian}
\quad  R_{ \{1 \} }(\Delta) = R_{ \{2 \} }(\Delta)  
  & = \frac{1}{2} \log \frac{ \sigma^2(1 + \mathbbm{E}^2[r_{\theta}] ) }{ 
 {\Delta - \sigma^2(1 - \mathbbm{E}^2 [ r_{\theta} ]) }
  },
 \quad  \sigma^2(1 - \mathbbm{E}^2 [ r_{\theta}  ]) \leq \Delta 
 \leq 2 \sigma^2.
\end{align}
Evaluating \eqref{eq:USRDf_nonBayesian}, the nonBayesian USRDf is
\begin{align} \label{eq:ex3_nonbayesian}
 R_{ \{1 \} }(\Delta) = R_{ \{2 \} }(\Delta)  = \frac{1}{2} \log \frac{ \sigma^2(1 + r_{\min}^2) }{ 
  {\Delta - \sigma^2(1 - r_{\min}^2 ) }    },
 \qquad  \sigma^2(1 -  r_{\min}^2)  \leq \Delta 
 \leq 2 \sigma^2.
\end{align}
A simple comparison of \eqref{eq:ex3_bayesian} and \eqref{eq:ex3_nonbayesian}
shows that the nonBayesian USRDf is strictly larger than its Bayesian 
counterpart.
Also, it is seen from \eqref{eq:ex3_bayesian} and \eqref{eq:ex3_nonbayesian} above that
when $r_{\tau} > 0$  for all $ \tau \in \Theta$, 
the average correlation, $\mathbbm{E}[r_{\theta}]$,
and the smallest correlation,  $ r_{\min}  $, 
play similar roles in the expressions for 
Bayesian and nonBayesian USRDf, respectively.

\qed

\end{example}
 
 Lastly, the standard properties of the SRDf and the USRDf for GMMS
 and GMF  with fixed-set samplers are summarized in the lemma below,
 with the proof provided in Appendix \ref{app:convexity_std}. 
 
\begin{lemma} \label{l:convexity_std}
 The right-sides of \eqref{eq:SRDf_GMMS}, \eqref{eq:SRDf-gauss-field},
 \eqref{eq:USRDf_Bayesian} and \eqref{eq:USRDf_nonBayesian} are finite-valued, 
  decreasing, convex, continuous functions of 
  $\Delta_{\min,A} < \Delta \leq \Delta_{\max}$.
\end{lemma}

%%%%%%%%%%%%%%%%%%%%%%%%%%%%%%%%%%%%%%%%%%%%%

%%%%%%%%%%%%%%%%%%%%%%%%%%%%%%%%%%%%%%%%%%%%%

\section{Proofs} \label{s:Proofs}

\subsection{Achievability proofs}

We present first the achievability proof of Theorem \ref{th:GMMS_SRDf} where
the sampled components of the GMMS are reconstructed first with a weighted MSE distortion measure 
under a reduced threshold, and then MMSE estimates are formed for the unsampled
components based on the former.
An achievability proof for Proposition \ref{prop:SRDf_GMF} is along
similar lines. Building on this, we present next an
achievability proof for Theorem \ref{th:GMMS_USRDf} with an emphasis on the Bayesian setting.
All our achievability proofs emphasize the modular structure of the reconstruction
mechanism, which allows GMMS reconstruction to be performed in two steps.

\vspace*{0.1cm}
 
\noindent {\bf Theorem \ref{th:GMMS_SRDf}}: First, observe that 
\begin{align}
\quad \Delta_{\min,A} 
& = \underset{ X_{A^c} \MC X_{A} \MC Y_{\cM} } \min \mathbbm{E}[||X_{\cM}- Y_{\cM}||^2] \\
& = \underset{ X_{A^c} \MC X_{A} \MC Y_{A^c} } \min 
\sum \limits_{i \in A^c} \mathbbm{E}[ (X_i - Y_i)^2  ] \qquad \text{with } Y_i = X_i, \ i \in A \\
& = \sum \limits_{i \in A^c} \mathbbm{E}[ (X_i -  \mathbbm{E}[X_i|X_A] )^2  ] \\
& = \sum \limits_{i \in A^c} \big (  \mathbbm{E}[X_i^2]  - \mathbbm{E}[X_i X_{A}^T] {\bf \Sigma}_{A}^{-1}
\mathbbm{E}[X_A X_i] \big )
\end{align}
and 
\begin{align}
 \Delta_{\max}
 & = \underset{ X_{A^c} \MC X_{A} \MC Y_{\cM} \atop X_{A} \independent Y_{\cM} } \min 
 \mathbbm{E}[ || X_{\cM} - Y_{\cM}||^2] \\
 & = \underset{ y_{\cM} } \min \ \mathbbm{E}[ || X_{\cM} - y_{\cM}||^2] \\
 & = \sum \limits_{i=1}^{m} \mathbbm{E}[X_i^2],
\end{align}
where ${\bf \Sigma}_{A}^{-1}$ exists by the assumed positive-definiteness
of ${\bf \Sigma}_{\cM}.$

 \vspace*{0.1cm}
 
Given $\ep>0$, for the GMMS $\{ X_{A t} \}_{t=1}^{\infty}$ 
 with pdf $\cN( {\bf 0}, {\bf \Sigma}_{A})$
 and weighted MSE distortion measure $d_A$, 
 consider a (standard) rate distortion
 code $(f_A, \varphi_A), \ f_A : \mathbbm{R}^{nk} \rightarrow    \{1, \ldots, J \} $ 
 and 
 $\varphi_A: \{1, \ldots, J \} \rightarrow \mathbbm{R}^{nk}$ of rate 
 $\frac{1}{n} \log J \leq \rho_{A}(\Delta) + \ep$
and with
\begin{align}
\mathbbm{E}[d_{A}(X_{A}^{n}, Y_{A}^{n})] \leq \Delta - \Delta_{ {\min},A} + \ep, 
\end{align}
for $n \geq N_{\ep}$, say.

\vspace*{0.2cm}

\noindent A code $(f,\varphi)$ is devised as follows. The encoder 
$f$ is chosen to be $f_A$, i.e.,
\begin{align}
f(x_A^n) \triangleq f_A(x_A^n), \quad x_A^n \in \mathbbm{R}^{nk}
\end{align}
and the decoder $\varphi$ is given by
\begin{align} \label{eq:ach_SRD_code}
\quad \varphi(j) \triangleq \big (  \varphi_A(j), \ 
\mathbbm{E}[X_{A^c}^n|X_A^n = \varphi_A(j)] \big ), \quad 
j \in \{1, \ldots, J \}.
\end{align}
The rate of the code $(f, \varphi)$ is 
$$\frac{1}{n} \log J \leq \rho_{A}(\Delta) + \ep.$$ 
Denote the output of the decoder $\varphi(f(X_{A}^n))$
by $Y_{\cM}^n = (Y_{A}^n, Y_{A^c}^n)$. Then, 
$Y_{A^c}^n = {\bf \Sigma}_{A^c A} {\bf \Sigma}_{A} Y_{A}^n$
and by the standard properties of an MMSE estimate,
for $t = 1, \ldots,n,$ it holds that
\begin{align} \label{eq:SRDf_ortho}
(X_{A^c t} - {\bf \Sigma}_{A^c A} {\bf \Sigma}_{A}^{-1} X_{A t}) \independent X_{A t}.
\end{align}
The code $(f , \varphi)$ has expected distortion
\begin{align}
 \mathbbm{E}[||X_{\cM}^n - Y_{\cM}^n ||^2 ]
 & = \mathbbm{E}[ || X_{A}^n - Y_{A}^n ||^2 ] + 
 \mathbbm{E}[ || X_{A^c}^n -   Y_{A^c}^n ||^2 ] \label{eq:SRDf_ach_eq3} \\
 & = \mathbbm{E}[ || X_{A}^n - Y_{A}^n ||^2 ] + 
 \mathbbm{E}[ || X_{A^c}^n -  {\bf \Sigma}_{A^c A} {\bf \Sigma}_{A}^{-1} Y_{A}^n ||^2 ] \\
 & = \mathbbm{E}[ || X_{A}^n - Y_{A}^n ||^2 ]
  + \mathbbm{E}[ || X_{A^c}^n - {\bf \Sigma}_{A^c A} {\bf \Sigma}_{A}^{-1} X_{A}^n 
   + {\bf \Sigma}_{A^c A} {\bf \Sigma}_{A}^{-1} X_A^n -  {\bf \Sigma}_{A^c A} {\bf \Sigma}_{A}^{-1} Y_{A }^n ||^2 ] \\
 & =  \mathbbm{E}[ || X_{A}^n - Y_{A}^n ||^2 ] 
  +  \mathbbm{E}[ || X_{A^c}^n - {\bf \Sigma}_{A^c A} {\bf \Sigma}_{A}^{-1} X_{A}^n ||^2 ]
  + \mathbbm{E}[ || {\bf \Sigma}_{A^c A} {\bf \Sigma}_{A}^{-1}  X_A^n -
   {\bf \Sigma}_{A^c A} {\bf \Sigma}_{A}^{-1}Y_{A}^n ||^2 ] \label{eq:SRDf_ach_eq1}\\
 & =  \Delta_{\min,A} + \frac{1}{n} \sum \limits_{t=1}^{n}
 \mathbbm{E}[ ( X_{A t} - Y_{A t} )^T(I + {\bf \Sigma}_{A}^{-1} {\bf \Sigma}_{A A^{c}} 
 {\bf \Sigma}_{A^c A}  {\bf \Sigma}_{A}^{-1} ) (X_{A t} - Y_{A t})  ] \label{eq:SRDf_ach_eq6}  \\
 & = \Delta_{\min,A} + \mathbbm{E}[d_{A}(X_{A}^n, Y_{A}^n) ] \\ 
 & \leq \Delta + \ep, \label{eq:SRDf_ach_eq4}
\end{align}
where  \eqref{eq:SRDf_ach_eq1} is by the orthogonality principle 
of MMSE estimates \eqref{eq:SRDf_ortho} and since
for $t = 1, \ldots, n$,
\begin{align}
 \mathbbm{E}[(X_{A^c t} - {\bf \Sigma}_{A^c A} {\bf \Sigma}_{A}^{-1} X_{A t})^T 
 {\bf \Sigma}_{A^c A}  {\bf \Sigma}_{A}^{-1} Y_{A t}]
 \! & =  \mathbbm{E}[ X_{A^c t} ^T 
 {\bf \Sigma}_{A^c A}  {\bf \Sigma}_{A}^{-1} Y_{A t}]
 -  \mathbbm{E}[ X_{A t}^T {\bf \Sigma}_{A}^{-1}   {\bf \Sigma}_{A A^{c}}  
 {\bf \Sigma}_{A^c A}  {\bf \Sigma}_{A}^{-1} Y_{A t}] \\
 & = \mathbbm{E} \big[ \mathbbm{E}[ X_{A^c t} ^T 
 {\bf \Sigma}_{A^c A}  {\bf \Sigma}_{A}^{-1} Y_{A t}|X_{A}^n] \big]
 \! - \!  \mathbbm{E}[ X_{A t}^T {\bf \Sigma}_{A}^{-1}   {\bf \Sigma}_{A A^{c}}  
 {\bf \Sigma}_{A^c A}  {\bf \Sigma}_{A}^{-1} Y_{A t}] \\
 & = \mathbbm{E} \big[ \mathbbm{E}[ X_{A^c t} ^T 
 |X_{A}^n] {\bf \Sigma}_{A^c A}  {\bf \Sigma}_{A}^{-1} Y_{A t} \big]
 -  \mathbbm{E}[ X_{A t}^T {\bf \Sigma}_{A}^{-1}   {\bf \Sigma}_{A A^{c}}  
 {\bf \Sigma}_{A^c A}  {\bf \Sigma}_{A}^{-1} Y_{A t}] \\
 & = \mathbbm{E}[ X_{A t}^T {\bf \Sigma}_{A}^{-1}   {\bf \Sigma}_{A A^{c}}  
 {\bf \Sigma}_{A^c A}  {\bf \Sigma}_{A}^{-1} Y_{A t}] - \mathbbm{E}[ X_{A t}^T {\bf \Sigma}_{A}^{-1}   {\bf \Sigma}_{A A^{c}}  
 {\bf \Sigma}_{A^c A}  {\bf \Sigma}_{A}^{-1} Y_{A t}] \\
 & = 0.
\end{align}

\qed

\vspace*{0.1cm}

\noindent {\bf Proposition \ref{prop:SRDf_GMF}}: The achievability proof 
of Proposition \ref{prop:SRDf_GMF} is along the lines of 
Theorem \ref{th:GMMS_SRDf}. For a given $\Delta_{\min} < \Delta \leq \Delta_{\max}$
and $\ep > 0$, for the GMMS $\{X_{A t} \}_{t=1}^{\infty}$ with 
weighted MSE distortion measure 
$$d_A(x_A, y_A) \triangleq ( x_A - y_A )^T {\bf G}_{A, I} (x_A - y_A),
 \ \ x_A, y_A \in \mathbbm{R}^k,
$$
consider a rate distortion code 
$(f_A, \varphi_A), \ f_A : \mathbbm{R}^{nk} \rightarrow    \{1, \ldots, J \} $ 
 and 
 $\varphi_A: \{1, \ldots, J \} \rightarrow \mathbbm{R}^{nk}$ of rate 
 $\frac{1}{n} \log J \leq \rho_{A}(\Delta) + \ep$
and with
\begin{align}
\mathbbm{E}[d_{A}(X_{A}^{n}, Y_{A}^{n})] \leq \Delta - \Delta_{ {\min},A} + \ep, 
\end{align}
for $n \geq N_{\ep}$.

\vspace*{0.2cm}

A code $(f, \varphi)$ is then constructed as follows. The encoder $f$ is 
chosen to be 
\begin{align}
 f(x_A^n) = f_{A}(x_{A}^n), \qquad x_{A}^n \in \mathbbm{R}^{nk}.
\end{align}
The output of decoder $\varphi$, corresponding to each $u \in I$, 
is given by 
\begin{align}
 (\varphi(j))_{u}  = \mathbbm{E}[X_{u}^n| X_{A}^n = \varphi_{A}(j) ],
 \qquad j \in \{1, \ldots, J \}.
\end{align}
Denoting the output of the decoder $\varphi(f(X_{A}^n))$ by $Y_{I}^n$,
for $ u \in I  , \  t = 1, \ldots,n,$
\begin{align}
 Y_{u t} = {\bf \Sigma}_{ \{u\} A} {\bf \Sigma}_{A}^{-1} Y_{A t}, 
\end{align}
where ${\bf \Sigma}_{ \{u\} A} = \mathbbm{E}[X_{u} X_{A}^T]$ and 
${\bf \Sigma}_{A} = \mathbbm{E}[X_{A} X_{A}^T]$.
The rate of the code $(f, \varphi)$ is 
$$\frac{1}{n} \log J \leq \rho_{A} (\Delta) + \ep.$$
The code $(f, \varphi)$ has expected distortion 
\begin{align}
 \mathbbm{E}[|| X_{I}^n - Y_{I}^n  ||^2]
 &   =  \int \limits_{I} 
       \mathbbm{E} \big[ || X_{u }^n - Y_{u}^n ||^2 \big] \, du \\ 
 & = \int \limits_{I  }  
       \mathbbm{E} [ || X_{u}^n -  {\bf \Sigma}_{ \{u\} A} {\bf \Sigma}_{A}^{-1} Y_{A}^n ||^2] \, du       
       \\
& =   \int \limits_{I }  
       \mathbbm{E} \big [ || X_{u}^n -  {\bf \Sigma}_{ \{u\} A} {\bf \Sigma}_{A}^{-1} X_{A}^n
       + {\bf \Sigma}_{ \{u\} A} {\bf \Sigma}_{A}^{-1} X_{A}^n - {\bf \Sigma}_{ \{u\} A} {\bf \Sigma}_{A}^{-1} Y_{A}^n ||^2 \big ] \, du \label{eq:GMF_ach_eq1} \\             
 & =  \int \limits_{I}  
       \mathbbm{E} \big [ || X_{u}^n -  {\bf \Sigma}_{ \{u\} A} {\bf \Sigma}_{A}^{-1} X_{A}^n ||^2 \big]
       + \mathbbm{E} \big [ ||{\bf \Sigma}_{ \{u\} A} {\bf \Sigma}_{A}^{-1} X_{A}^n -  {\bf \Sigma}_{ \{u\} A} {\bf \Sigma}_{A}^{-1} Y_{A}^n ||^2 \big] \, du \ \ \  \label{eq:GMF_ach_eq_ortho} \\
  & =  \Delta_{\min,A} 
       +   \int \limits_{I} 
       \mathbbm{E} \big[ || (X_{A}^n -  Y_{A}^n)^T {\bf \Sigma}_{A}^{-1} {\bf \Sigma}_{ \{u\} A}^T
        {\bf \Sigma}_{ \{u\} A} {\bf \Sigma}_{A}^{-1} (X_{A}^n -  Y_{A}^n) ||^2   \big ] \, du \\
  & =    \Delta_{\min,A} + 
         \mathbbm{E} \big[ || (X_{A}^n -  Y_{A}^n)^T {\bf G}_{A, I}  (X_{A}^n -  Y_{A}^n) ||^2 \big] 
         \qquad \text{ by } \eqref{eq:GMF_G}
         \\
  & \leq \Delta + \ep,       
\end{align}
where \eqref{eq:GMF_ach_eq_ortho} is by the orthogonality principle 
of the MMSE estimates as in \eqref{eq:SRDf_ach_eq1}, \eqref{eq:SRDf_ach_eq6}.

\qed

\vspace*{0.1cm}

%%%%%%%%%%%%%%%%%%%%%%

Before we present the achievability proof of Theorem \ref{th:GMMS_USRDf}, 
we present pertinent technical results. 
We state first a standard technical result, a Vitali covering lemma
(Theorem 17.1 in \cite{DiBenedetto02}), without proof.
For any   $\ep > 0$, this lemma guarantees the existence of a 
finite number of nonoverlapping Euclidean ``balls'' of radius $\leq \ep$
such that the Lebesgue measure of the set of members of $\Theta_1$ 
not covered by the Euclidean balls is $\leq \ep$.
In the achievability proof of Theorem \ref{th:GMMS_USRDf},
the centers of such balls will be used to approximate $\Theta_1$
and (approximately) estimate  $\theta_1$.
For $\tau_1 \in \Theta_1$, let $B_{\tau_1, \ep} \subset \mathbbm{R}^{k^2}$
denote a standard Euclidean $\ell_2$-ball with center  $\tau_1$ and radius $\ep$.

\begin{lemma} \label{l:covering}
For every $\ep > 0$, there exists an $N_{\ep} > 0$ and a finite 
disjoint collection of balls $\{ B_{ \tau_{ 1,i}, \ep_i } \}_{i=1}^{N_{\ep}}$
such that $\underset{i} \max \ \ep_i \leq \ep$ and
\begin{align} \label{eq:excess_cover}
 \mu \big ( \Theta_1 \setminus \underset{i} \bigcup  B_{ \tau_{ _{1,i}}, \ep_i } \big )
 < \ep,
\end{align}
where $ \mu $ is the Lebesgue measure on $\mathbbm{R}^{k^2}$
and $\setminus $ is the standard set difference.
\end{lemma}  
 
\noindent {\it Remarks}: i) The lemma above relies on $\Theta_1$
being a compact subset of $\mathbbm{R}^{k^2}.$

\vspace*{0.1cm}

\noindent ii) For $\ep > 0$  and 
$\{ B_{ \tau_{ _{1,i}}, \ep_i } \}_{i=1}^{N_{\ep}}$ as in the lemma above, 
let $\Theta_{ 1,\ep} \subset \Theta_1 $  
be the collection of ``centers''  $\{ \tau_{1,i} \}_{i=1}^{N_{\ep}}$.

%%%%%%%%%%%%%%%%%%%%%%

\vspace*{0.1cm}
 
While the lemma above is pertinent to the Bayesian and nonBayesian
parts of Theorem \ref{th:GMMS_USRDf},
Lemmas \ref{l:cont_condnl_dist} and \ref{l:part_close} below are 
pertinent to the Bayesian and nonBayesian settings respectively.

\begin{lemma} \label{l:cont_condnl_dist}
In the Bayesian setting, for every $x_{\cM} \in \mathbbm{R}^{m}$,
\begin{align} \label{eq:cont_condnl_dist}
\nu_{X_{\cM}|\theta_1}(x_{\cM}| \tau_1) 
\end{align}
is continuous in $\tau_1$. For any 
code $(f, \varphi), \ f : \mathbbm{R}^{nk} \rightarrow \{1, \ldots, J \},
\ \ \varphi : \{1, \ldots, J \} \rightarrow \mathbbm{R}^{nm}  $,
\begin{align}
\mathbbm{E}\big [ || X_{\cM}^n -  \varphi(f(X_A^n)) ||^2 \big |\theta_1 = \tau_1\big] 
\end{align}
is continuous in $\tau_1$.
\end{lemma}

\noindent {\it Proof}: See Appendix \ref{app:Bayesian}. \qed

\vspace*{0.2cm}

\noindent {\it Remarks}: (i) 
Since $\Theta_1$ is a compact set, 
for every $x_{\cM} \in \mathbbm{R}^m$, 
the pdf $\nu_{X_{\cM}|\theta_1}(x_{\cM}| \tau_1)$
and $\mathbbm{E}\big [ || X_{\cM}^n - \varphi(f(X_A^n)) ||^2 \big |\theta_1 = \tau_1\big]  $
are, in fact, uniformly continuous in $\tau_1$. 
Thus, for every $x_{\cM} \in \mathbbm{R}^m $ and $ \ep > 0$, 
there exists a $\delta > 0$ such that 
for $\tau_{1,1}, \ \tau_{1,2} \in \Theta_1$ with 
$ ||  \tau_{1,1} - \tau_{1,2} || \leq \delta$,
it holds that
\begin{align}
 | \nu_{X_{\cM}|\theta_1}(x_{\cM}| \tau_{1,1}) - \nu_{X_{\cM}|\theta_1}(x_{\cM}| \tau_{1,2}) | \leq \ep, 
\end{align}
and
\begin{align}
 \Big | \mathbbm{E}\big [ || X_{\cM}^n - \varphi(f(X_A^n)) ||^2 \big |\theta_1 = \tau_{1,1} \big]
 - \mathbbm{E}\big [ || X_{\cM}^n - \varphi(f(X_A^n)) ||^2 \big |\theta_1 = \tau_{1,2}\big]
 \Big | \leq \ep.
\end{align}

\noindent (ii) The claim \eqref{eq:cont_condnl_dist} implies that 
$$\mathbbm{E}[X_{A} X_{A}^T|\theta_1 = \tau_1] \quad 
\text{ and } \quad 
\mathbbm{E}[X_{A} X_{A^c}^T|\theta_1 = \tau_1]
$$ 
are continuous in $\tau_1$ and hence,
$${\bf G}_{A, \tau_1} = {\bf I} + (\mathbbm{E}[X_{A} X_{A}^T|\theta_1 = \tau_1])^{-1}
\mathbbm{E}[X_{A} X_{A^c}^T|\theta_1 = \tau_1]
\mathbbm{E}[X_{A^c} X_{A }^T|\theta_1 = \tau_1]
(\mathbbm{E}[X_{A} X_{A}^T|\theta_1 = \tau_1])^{-1}
$$
is continuous in $\tau_1$. Thus, from \eqref{eq:Bay_prim_equiv_form},
for every $\delta > \Delta_{\min,A, \tau_1}$,
$\rho_A^{\cB}(\delta, \tau_1)$ is continuous in $\tau_1$.

\vspace*{0.2cm}

\noindent The following lemma implies that if $\tau_{1,1}, \tau_{1,2} \in \Theta_1$
are ``close,'' then there exist ${\hat \tau}$ and $ {\check \tau}  $ in the ambiguity
atoms of $\tau_{1,1}$ and $\tau_{1,2}$, respectively, which too are ``close.''

\begin{lemma} \label{l:part_close}
 For every $\ep > 0$, there exists a $\delta > 0$ such that 
 for every $\tau_{1,1}, \tau_{1,2} \in \Theta_1$  with 
 $|| {\tau}_{1,1} - \tau_{1,2} || \leq \delta $, it holds that
 \begin{align} \label{eq:close_def_theta_1}
 \underset{\hat \tau  \in \Lambda(\tau_{1,1} ), \  \check \tau  \in \Lambda(\tau_{1,2} )} \min \
  || \hat \tau - \check \tau || \leq \ep.
 \end{align}
\end{lemma} 
 
\noindent {\it Proof}: See Appendix \ref{app:part_close}. 
\qed

%%%%%%%%%%%%%%%%%%%%%%

\vspace*{0.2cm}

\noindent {\bf Theorem \ref{th:GMMS_USRDf}}: 
Consider $\Theta_1$  as in Section \ref{s:Results}. 
Based on the output of the fixed-set sampler $X_A^n$, the encoder 
forms a {\it maximum-likelihood} (ML) estimate for the covariance-matrix ${\bf \Sigma}_{A \tau_1}$
as
\begin{align} \label{eq:ML_est_def} 
{\widehat \theta}_{1,n} = {\widehat \theta}_{1,n}(X_A^n) = 
\frac{1}{n} \sum \limits_{t=1}^{n} X_{A t} X_{A t}^T .
\end{align}
Observe that $\{ X_{At} \}_{t=1}^{\infty}$ is a GMMS with pdf 
${\cal N}( {\bf 0}, {\bf \Sigma}_{A \tau_1})$
and $\nu_{X_{A}|\theta_1 = \tau_1}$ is continuous in
$\tau_1$.
Compactness of $\Theta_1$, the boundedness and continuity of 
$\nu_{X_A|\theta_1 = \tau_1}$ in $\tau_1$ imply, by the 
law of large numbers \cite{Levy}, that 
\begin{align}
{\widehat \theta}_{1,n} \overset{ a.s. } \longrightarrow  \tau_1,
\end{align}
where the convergence is elementwise and is
{under $ \nu_{X_{A}|\theta_1 = \tau_1}$},
and that for every
$\ep_1 > 0$, there exists a $\delta$ and $N_{\ep_1}$ such 
that for every $\tau_1 \in \Theta_1$
\begin{align} \label{eq:ML_est_Bay}
 P_{\tau_1} \big ( || \tau_1 - {\widehat \theta}_{1,n} || > \delta \big )
 \leq \ep_1, \quad n \geq N_{\ep_1}.
\end{align}
Now, considering a subset $\Theta_{1, \delta }$ of $\Theta_1$ as in the 
remark following Lemma \ref{l:covering},
 define ${\tilde \theta}_{1,n}$  as 
\begin{align} \label{eq:ML_cove_Bay}
 {\tilde \theta}_{1,n} \triangleq \underset{ {\check \tau_1} \in \Theta_{1, \delta}} {\arg \min}
 \ || {\widehat \theta}_{1,n} - {\check \tau_1}  ||.
\end{align}
Fixing $\ep > 0$ and $0 < \ep_1 < \ep$,
from \eqref{eq:ML_est_Bay}, \eqref{eq:ML_cove_Bay} and Lemma \ref{l:covering}, it follows that
there exists a $\delta$ and $N_{\ep_1}$ such that 
\begin{align} \label{eq:ML_close_Bay}
 P \big ( || \theta_1 - {\tilde \theta}_{1,n} || > 2 \delta \big )
 \leq \ep_1, \quad n \geq N_{\ep_1}.
\end{align}
Notice that while ${\hat \theta}_{1,n}$ may lie outside $\Theta_1$,
${\tilde \theta}_{1,n} $ is an estimate of $\theta_1$ that takes values 
in a finite subset of $\Theta_1$. 
 The estimate ${\tilde \theta}_{1,n} $ (of $\theta_1$) will be used in the 
 next part of the proof to select sampling rate distortion codes.

\vspace*{0.3cm}

For a fixed $\Delta_{\min, A} < \Delta \leq \Delta_{\max}$, let 
$\{ \Delta_{\tau_1}, \tau_1 \in \Theta_1 \}$ be such that it
attains the minimum in \eqref{eq:USRDf_Bayesian} and 
$\Delta_{\tau_1}$ is continuous in $\tau_1$ (see the remark below 
Theorem \ref{th:GMMS_USRDf}). 
Recall that for each $\tau_1 \in \Theta_{1, \delta}$, 
$\rho_A^{\cB}(\Delta_{\tau_1}, \tau_1) $ is, in effect,
the RDf for a GMMS $\{X_{A t} \}_{t=1}^{\infty}$
with pdf $\nu_{X_A|\theta_1 = \tau_1}$ under a weighted MSE distortion
measure $d_{A \tau_1}$.
Thus, for each $\tau_1 \in \Theta_{1, \delta}$, 
there exists a (standard) rate distortion code $(f_{\tau_1}, \varphi_{\tau_1}), \ 
f_{\tau_1}: \mathbbm{R}^{nk} \rightarrow \{1, \ldots, J \}$ and 
$\varphi_{\tau_1} : \{1,\ldots, J \} \rightarrow  \mathbbm{R}^{nk}$
of rate $\frac{1}{n} \log J \leq \rho_A^{\cB}(\Delta_{\tau_1}, \tau_1) 
+ \ep_1 \leq R_{A}(\Delta) + \ep_1$ 
and with 
\begin{align}
 \mathbbm{E}[d_{A \tau_1}(X_A^n, \varphi_{\tau_1}(f_{\tau_1}(X_{A}^n))) | \theta_1 = \tau_1]
 \leq \Delta_{\tau_1} - \Delta_{\min, A,\tau_1 } + \ep_1
\end{align}
for all $n \geq N_{\ep_1}$.

\vspace*{0.2cm}

Now, consider a code $(f, \varphi)$ with $f$ taking values in
$\cJ \triangleq \{1, \ldots, |\Theta_{1, \delta}| \}
\times \{1, \ldots, J \}$  as follows. 
Order (in any manner) the elements of $\Theta_{1, \delta}$.
The encoder $f$, dictated by the estimate ${\tilde \theta}_{1,n}$, is given by
\begin{align} \label{eq:Bay_encoder}
 f(x_A^n) \triangleq \big ( {\tilde \theta}_{1,n}(x_{A}^n), f_{{\tilde \theta}_{1,n} }(x_A^n) \big), 
 \quad x_A^n \in \mathbbm{R}^{nk}  \quad 
\end{align}
and the decoder $\varphi$ is given by
\begin{align} \label{eq:Bay_decoder}
\quad  \varphi( \tau_1 , j ) \triangleq  \big ( \varphi_{\tau_1}(j),
  \mathbbm{E}[ X_{A^c}^n | X_A^n = \varphi_{\tau_1}(j), \theta_1 = \tau_1 ] \big ), 
  \quad  (\tau_1,j) \in \cJ. 
\end{align}

\noindent By the finiteness of $\Theta_{1, \delta },$ the rate of the code $(f, \varphi)$ is 
\begin{align}
 \frac{1}{n} \log |\cJ| 
 & = \frac{1}{n} \log |\Theta_{1, \delta}| + \frac{1}{n} \log J \\
 & \leq R_{A} (\Delta) + 2 \ep_1,
\end{align}
for $n$ large enough. 
Denoting the output of the decoder by $Y_{\cM}^n$ with 
$Y_A^n = \varphi_{\tau_1}(j)$ and $Y_{A^c}^n = 
\mathbbm{E}[ X_{A^c}^n | X_A^n = \varphi_{\tau_1}(j), \theta_1 = \tau_1 ]$,
we have that
\begin{align} 
 \mathbbm{E}[ || X_{\cM}^n - Y_{\cM}^n ||^2 ]
 & \! = \! \mathbbm{E}[ \mathbbm{1}( ||  {\tilde \theta}_{1,n} - \theta_1 || \leq 2 \delta ) 
 || X_{\cM}^n - Y_{\cM}^n ||^2 ] 
     \! +  \! \mathbbm{E}[ \mathbbm{1}( ||  {\tilde \theta}_{1,n} - \theta_1 || > 2 \delta ) 
  || X_{\cM}^n - Y_{\cM}^n ||^2 ] . 
 \label{eq:Distortion_Bay_USRDf}
\end{align}
Using Lemma \ref{l:cont_condnl_dist}, it is shown in Appendix 
\ref{app:Bay_correct_est_dist} that 
the first term in the right-side 
of \eqref{eq:Distortion_Bay_USRDf} is
\begin{align}
 \mathbbm{E}[ \mathbbm{1}( ||  {\tilde \theta}_{1,n} - \theta_1 || \leq 2 \delta ) 
  || X_{\cM}^n - Y_{\cM}^n ||^2 ] \leq \Delta + 4 \ep_1.
 \label{eq:Bay_correct_est_dist}
\end{align}

\vspace*{0.2cm}

Next, we show that the second term in the right-side of 
\eqref{eq:Distortion_Bay_USRDf} is ``small.''
First, note that the finiteness of $ \Theta_{1,\delta}$ implies the 
existence of an $M_1$ such that, for $t=1,\ldots,n,$
$$ | (\varphi_{\tau_1}(f_{\tau_1}(x_A^n) ))_{i,t} | \leq M_1,  \ i \in A,  
\ \tau_1 \in \Theta_{1, \delta}, \ \ \  x_A^n \in \mathbbm{R}^{nk} $$ 
and hence, from \eqref{eq:Bay_decoder}, there exists an
$M_2 > 0$ such that, for $t = 1, \ldots,n,$
\begin{align} \label{eq:Bound_Y_Bay}
  | (\varphi(f(x_A^n) ))_{i,t} | \leq M_2 , \quad 
 i \in \cM, \  x_A^n \in \mathbbm{R}^{nk}.
\end{align}
For $i \in \cM$, from \eqref{eq:Bound_Y_Bay}, Cauchy-Schwarz inequality,
and the fact that $\mathbbm{E}[X_i^2]$ is bounded, 
%$\Theta$ is compact,
there exists an $M$ such that
\begin{align} \label{eq:Bound_code_Bay}
\qquad \mathbbm{E}[ (X_{i t}  -Y_{i t})^4 ] \leq M, \qquad t = 1, \ldots,n.
\end{align}
Now, the second term on the right-side of \eqref{eq:Distortion_Bay_USRDf},
\begin{align}
\ \  \mathbbm{E}[ \mathbbm{1}( ||  {\tilde \theta}_{1,n} - \theta_1 || > 2 \delta )
    || X_{\cM}^n - Y_{\cM}^n ||^2 ]
 & = \frac{1}{n} \sum \limits_{t=1}^n \sum \limits_{i=1}^m 
 \mathbbm{E} \big [ \mathbbm{1}( ||  {\tilde \theta}_{1,n} - \theta_1 || > 2 \delta )
 \big (X_{i t} - Y_{i t} \big)^2 \big ] \\
 & \leq \frac{1}{n} \sum \limits_{t=1}^n \sum \limits_{i=1}^m 
        \sqrt{ \mathbbm{E} \big [ \mathbbm{1}^2( 
        ||  {\tilde \theta}_{1,n} - \theta_1 || > 2 \delta )  \big ] 
        \mathbbm{E} \big [  \big (X_{i t} - Y_{i t} \big)^4 \big ] }
    \label{eq:cauchy_schwarz}    \\
 & \leq \frac{1}{n} \sum \limits_{t=1}^n \sum \limits_{i=1}^m \sqrt{  \ep_1 
 M } \hspace*{2.7 cm} \text{ from } \eqref{eq:ML_close_Bay}
 \text{ and } \eqref{eq:Bound_code_Bay}
 \\
 & \leq \sqrt{\ep_1 M} m,     \label{eq:Bay_wrong_est_dist} 
\end{align}
where \eqref{eq:cauchy_schwarz} is by the Cauchy-Schwarz inequality.
From \eqref{eq:Bay_correct_est_dist} and 
\eqref{eq:Bay_wrong_est_dist}, we get
\begin{align}
 \mathbbm{E} \big [  || X_{\cM}^n - Y_{\cM}^n ||^2 \big]
 & \leq \Delta + 4 \ep_1  + m \sqrt{\ep_1 M} \\
 & \leq \Delta + \ep,
\end{align}
for $\ep_1$ small enough.

\vspace*{0.3cm}

In the nonBayesian setting, as a first step, Lemma \ref{l:part_close} is used
to show that $\rho_{A}^{n \cB}(\delta, \tau_1)$ is a continuous function 
of $\tau_1$. Then, the maximum in \eqref{eq:USRDf_nonBayesian} is seen to exist as 
a continuous function over a compact set attains its supremum. Next,
the achievability proof follows by adapting the steps above
with the following differences.
For each $\tau_1 \in \Theta_{1,\delta},$ sampling rate distortion 
codes $(f_{\tau_1}, \varphi_{\tau_1}), \ 
f_{\tau_1} : \mathbbm{R}^{nk} \rightarrow \{1, \ldots, J \}, \
\varphi_{\tau_1} : \{1, \ldots, J \} \rightarrow \mathbbm{R}^{nm}$ 
are chosen to satisfy 
\begin{align}
 \mathbbm{E} \big [ || X_{\cM}^n - \varphi_{{\tau_1}} ( f_{\tau_1}(X_A^n)) ||^2 | \theta = \tau \big ]
 \leq \Delta , \quad \tau \in \Lambda(\tau_1),
\end{align}
with rate $\frac{1}{n} \log || f_{\tau_1} || \leq R_A(\Delta ) + \ep$, 
where $R_{A}(\Delta)$ is the nonBayesian USRDf. A code
$(f, \varphi)$ with $f$ taking values in 
$\cJ = \{1, \ldots, |\Theta_{1, \delta}| \}
\times \{1, \ldots, J \}$ is constructed based on the codes
$(f_{\tau_1}, \varphi_{\tau_1})$ as before. 
While counterparts of  \eqref{eq:Bay_correct_est_dist}
and \eqref{eq:Bay_wrong_est_dist} can be shown for each
$\tau_1 \in \Theta_1$ using a similar set of ideas,
a key distinction in the analysis is that Lemma \ref{l:part_close}
is used in lieu of Lemma \ref{l:cont_condnl_dist} to show that 
\begin{align}
\quad  \mathbbm{E} \big [ \mathbbm{1}( ||  {\tilde \theta}_{1,n} - \tau_1 || \leq 2 \ep_1 ) 
   || X_{\cM}^n - \varphi  ( {\tilde \theta}_{1,n}, 
   f_{{\tilde \theta}_{1,n}}(X_A^n) )  ||^2 \big | \theta  = \tau  \big]  
 & \leq \Delta + \ep_1, \quad \tau \in \Lambda(\tau_1), \ \tau_1 \in \Theta_1,
\end{align}
the counterpart of \eqref{eq:Bay_correct_est_dist}.
\qed

\subsection{Converse proof}

In contrast to the achievability proofs, we present a converse proof
for Theorem \ref{th:GMMS_USRDf} first, with an emphasis on the Bayesian setting; 
this is then adapted to Theorem \ref{th:GMMS_SRDf}.
Prior to this, we prove the equivalence of expressions in \eqref{eq:ach_fnl_equiv},
that will be pertinent to Theorem \ref{th:GMMS_SRDf}.
Building on this, we show the equivalence of the simplified forms for 
$\rho_A^{\cB}(\delta, \tau_1)$ and $\rho_A^{n\cB}(\delta, \tau_1)$
in Proposition \ref{prop:prim_USRDf_equiv_form}.
Next, we shall present a technical lemma. These will be used subsequently
in the unified converse proof for Theorems \ref{th:GMMS_SRDf} and \ref{th:GMMS_USRDf}.  
The converse proof for Proposition \ref{prop:SRDf_GMF} uses an approach
that does not rely on Lemma \ref{l:single_letter_markov2} and is presented last.

\vspace*{0.3cm}

\noindent {\bf Equivalence for Theorem \ref{th:GMMS_SRDf}}:
The following equality will be relevant in the proof of converse
for Theorem \ref{th:GMMS_SRDf}:
\begin{align} 
\underset{ \substack{X_{A^c} \MC X_A \MC Y_{\cM} \\ 
\mu_{X_A Y_{A}} < \hspace*{-0.1cm} < \mu_{X_{A}} \times \mu_{Y_{A}} \\
\mathbbm{E}[ || X_{\cM} - Y_{\cM}||^2]
\leq \Delta
} } \min I \big( X_{A} \wedge Y_{A} \big)
 = \displaystyle \min_{ \mu_{X_A Y_{A}} < \hspace*{-0.1cm} < \mu_{X_{A}} \times \mu_{Y_{A}} 
 \atop \mathbb{E} [ d_{A}(X_{A}, Y_{A}) ] \leq {\Delta} - {\Delta}_{ \min, A}} 
I \big( X_{A} \wedge Y_{A} \big). \label{eq:ach_fnl_equiv}
\end{align}
For any pair of rvs $X_{\cM}, Y_{\cM}$   
satisfying the constraints on the left-side of \eqref{eq:ach_fnl_equiv}, 
consider 
\begin{align}
 {\widehat Y}_{\cM} \triangleq \mathbbm{E}[X_{\cM}| Y_{\cM}]. 
 \label{eq:SRDf_GMMS_equiv_eq1}
\end{align}
Now,
\begin{align}
 {\widehat Y}_{A^c}  & =  \mathbbm{E}[X_{A^c}| Y_{\cM} ]
 = \mathbbm{E}[ \mathbbm{E}[X_{A^c}|X_A, Y_{\cM} ] | Y_{\cM}  ] 
  = \mathbbm{E}[ \mathbbm{E}[X_{A^c}|X_A ] | Y_{\cM}  ] 
  = \mathbbm{E}[ {\bf \Sigma}_{A^c A }{\bf \Sigma}_{A}^{-1}X_A | Y_{\cM}  ] 
  = {\bf \Sigma}_{A^c A } {\bf \Sigma}_{A}^{-1} {\widehat Y}_{A}.
  \label{eq:SRDf_GMMS_equiv_eq5}
\end{align}
By the optimality of the MMSE estimate, 
\begin{align}
\mathbbm{E}[ || X_{\cM} - {\widehat Y}_{\cM} ||^2   ] \leq 
\mathbbm{E}[ || X_{\cM} -  Y_{\cM} ||^2  ] \leq {\Delta }. \label{eq:SRDf_GMMS_equiv_eq3}
\end{align}
It is readily checked (along the lines of \eqref{eq:SRDf_ach_eq3}-\eqref{eq:SRDf_ach_eq4}) that 
\begin{align}
\mathbbm{E}[ || X_{\cM} - {\widehat Y}_{\cM} ||^2 ]
 = \mathbbm{E}[ d_{A} (X_{A}, {\widehat Y}_{A})  ] + \Delta_{\min, A  }.
 \label{eq:SRDf_GMMS_equiv_eq4}
\end{align}
Putting together \eqref{eq:SRDf_GMMS_equiv_eq1}-\eqref{eq:SRDf_GMMS_equiv_eq4},
completes the proof of  \eqref{eq:ach_fnl_equiv}.

\qed

\noeqref{eq:SRDf_GMMS_equiv_eq2} \noeqref{eq:SRDf_GMMS_equiv_eq3}
\noeqref{eq:SRDf_GMMS_equiv_eq5}
\vspace*{0.1cm}

\vspace*{0.1cm}

 \noindent {\bf Proposition \ref{prop:prim_USRDf_equiv_form}}: 
 The proof of \eqref{eq:Bay_prim_equiv_form} and \eqref{eq:nonBay_prim_equiv_form}
 is the along the lines of proof of \eqref{eq:ach_fnl_equiv}, with the distinction
 that in the nonBayesian setting, ${\widehat Y}_{A}$ is chosen to satisfy the 
 orthogonality principle and ${\widehat Y}_{A^c}$ is chosen to be a linear function
 of ${\widehat Y}_{A}$.

 \qed

\vspace*{0.2cm}

\noindent The following technical lemma is the counterpart of 
Lemma 6 in \cite{BodNar17U}.

\begin{lemma} \label{l:single_letter_markov2}
In the Bayesian setting, for any $n$-length $k$-FS code $(f, \varphi)$
with $ f: \mathbbm{R}^{nk} \rightarrow \{1, \ldots, J \}, \ 
\varphi: \{1, \ldots, J \} \rightarrow \mathbbm{R}^{nm}, \ $
for $t=1 , \ldots, n,$
denoting $\varphi(f(X_{A}^n))$ by $Y_{\cM}^n, $ 
it holds that
\begin{align} \label{eq:l_claim}
 \theta, X_{A^c t} \MC \theta_1, X_{A t} \MC Y_{\cM t}.
\end{align}
\end{lemma}

\noindent {\bf Proof}: First, note that 
\begin{align} \label{eq:l_s_block_markov}
 \theta, X_{A^c}^n \MC X_{A}^n \MC Y_{\cM}^n
\end{align}
holds by code construction.
From \eqref{eq:l_s_block_markov} (and since $Y_{\cM}^n$ above is a finite-valued
rv), we have
\begin{align}
 0  = I \big (\theta,X_{A^c}^n \wedge Y_{\cM}^n | X_{A}^n \big ) 
  & = I\big (\theta \wedge Y_{\cM}^n | X_{A}^n \big ) 
  + I \big (X_{A^c}^n \wedge Y_{\cM}^n | X_{A}^n, \theta \big )  \\
  & = I \big (\theta, \theta_1 \wedge Y_{\cM}^n | X_{A}^n \big )
    + I \big (X_{A^c}^n \wedge Y_{\cM}^n | X_{A}^n, \theta \big )  
   \label{eq:l_theta1_fn}\\
  & \geq I \big ( \theta \wedge Y_{\cM}^n | X_{A}^n, \theta_1 \big ) 
    + I \big (X_{A^c}^n \wedge Y_{\cM}^n | X_{A}^n, \theta \big ), 
    \label{eq:l_secondterm}
\end{align}
where \eqref{eq:l_theta1_fn}  is since $\theta_1$ is a function of $\theta.$
Now, the  second term on the right-side of \eqref{eq:l_secondterm} is
\begin{align}
 0 = I(X_{A^c}^n \wedge Y_{\cM}^n |X_{A}^n, \theta) 
 & = \sum \limits_{t=1}^n I \big(X_{A^c t} \wedge Y_{\cM}^n |X_{A^c}^{t-1}, X_{A}^n, \theta \big) \\
 & = \sum \limits_{t=1}^n I \big(X_{A^c t} \wedge X_{A^c}^{t-1}, X_{A}^{n \setminus t}, Y_{\cM}^n | X_{A t} , \theta \big) 
     - I \big(X_{A^c t} \wedge X_{A^c}^{t-1}X_{A}^{n \setminus t} | X_{A t}, \theta \big)  \\
 & =  \sum \limits_{t=1}^n I \big (X_{A^c t} \wedge X_{A^c}^{t-1}, X_{A}^{n \setminus t}, Y_{\cM}^n | X_{A t} , \theta \big ), 
 \qquad \text{since } \nu_{X_{\cM}^n|\theta} = \prod \limits_{t=1}^{n} \nu_{X_{\cM t}|\theta}  \\   
 & \geq \sum \limits_{t=1}^n I \big (X_{A^c t} \wedge  Y_{\cM t}  | X_{A t} , \theta \big )
 \label{eq:l_single_eq1} .
\end{align}
Next, \eqref{eq:l_secondterm} and the fact  
\begin{align} \label{eq:l_theta1_theta_indep}
 \theta \MC \theta_1 \MC X_{A}^n
\end{align}
imply 
\begin{align}
 0 & = I \big ( \theta \wedge  X_{A}^n | \theta_1 \big ) 
       + I \big ( \theta \wedge Y_{\cM}^n | X_{A}^n, \theta_1 \big )  \\
 & = I \big ( \theta \wedge X_{A}^n, Y_{\cM}^n | \theta_1 \big ) 
\end{align}
and hence, for $t = 1, \ldots, n,$
\begin{align}
 I(\theta \wedge  X_{A t}, Y_{\cM t} |\theta_1) = 0. \label{eq:l_single_eq2}
\end{align}
Now, by \eqref{eq:l_single_eq1} and \eqref{eq:l_single_eq2}, for $t = 1, \ldots,n,$
\begin{align}
 I(\theta \wedge Y_{\cM t} |X_{A t}, \theta_1) + I( X_{A^c t} \wedge Y_{\cM t} | X_{A t}, \theta ) = I(\theta, X_{A^c t} \wedge Y_{\cM t} |X_{A t}, \theta_1) = 0,
\end{align}
hence, the claim of the lemma \eqref{eq:l_claim}.

\qed

\vspace*{0.3cm}

\noindent {\bf Converse}: We provide first a converse 
proof for the Bayesian setting in Theorem \ref{th:GMMS_USRDf}, which is then 
refashioned to provide converse
proofs for the nonBayesian setting and Theorem \ref{th:GMMS_SRDf}.

\vspace*{0.1cm}

Let $(f, \varphi)$ be an $n$-length $k$-FS code of rate $R$ and with decoder
output $Y_{\cM}^n = \varphi(f(X_A^n))$ satisfying 
$\mathbbm{E}[||X_{\cM}^n - Y_{\cM}^n||^2] \leq \Delta$.
By lemma \ref{l:single_letter_markov2}, for $t = 1, \ldots,n,$ we have
\begin{align}
 \theta, X_{A^c t} \MC \theta_1, X_{A t} \MC Y_{\cM t}. 
 \label{eq:Conv_single_let_Markov}
\end{align}

\vspace*{0.1cm}

For $t=1, \ldots,n$, and $\tau_1 \in \Theta_1$, let 
$\Delta_{\tau_1, t}$ denote $\mathbbm{E}[||X_{\cM t} - Y_{\cM t}||^2|\theta_1 = \tau_1]$
and $\Delta_{\tau_1} \triangleq \frac{1}{n} \sum \limits_{t=1}^n 
\mathbbm{E}[ || X_{\cM t} - Y_{\cM t}||^2 |\theta_1 = \tau_1]$.
Along the lines of proof of Theorem 9.6.1 in \cite{Gallager}, for every $\tau_1 \in \Theta_1$,
\begin{align}
\quad  R = \frac{1}{n} \log | f | & \geq \frac{1}{n} H( f(X_A^n ) |\theta_1 = \tau_1) 
  \\
  & \geq  \frac{1}{n} H(Y_{A}^n | \theta_1 = \tau_1) \\
 & = \frac{1}{n} I(X_A^n \wedge Y_{A}^n | \theta_1 = \tau_1 )  \\ 
 & = \frac{1}{n} \sum \limits_{t=1}^n \Big ( 
 I \big ( X_{At} \wedge X_A^{t-1},Y_{A}^n | \theta_1 = \tau_1 \big)
      - I \big (X_{A t} \wedge X_{A}^{t-1}|\theta_1 = \tau_1 \big)  \Big ) 
      \\
 & = \frac{1}{n} \sum \limits_{t=1}^n 
      I(X_{A t} \wedge X_A^{t-1}, Y_{A}^n | \theta_1 = \tau_1) \hspace*{4.3cm}  
      \text{since } \ \ \nu_{X_{A}^n | \theta_1} = \prod \limits_{t=1}^{n} \nu_{X_{A t} | \theta_1}
      \\
 & \geq \frac{1}{n} \sum \limits_{t=1}^n 
         I(X_{A t} \wedge Y_{A t} | \theta_1 = \tau_1) \\
 & \geq \frac{1}{n} \sum \limits_{t=1}^n 
         \underset{ 
         \substack{\theta ,X_{A^c t} \MC \theta_1, X_{A t} \MC Y_{\cM t} \\ 
          \mu_{X_{A t} Y_{A t} | \theta_1= \tau_1} < \hspace*{-0.1cm} <  
          \mu_{X_{A t}   | \theta_1 = \tau_1} \times \mu_{ Y_{A t} | \theta_1= \tau_1} \\
         \mathbbm{E}[||X_{\cM t} - Y_{\cM t}||^2 |\theta_1 = \tau_1 ] \leq \Delta_{\tau_1,t}}}
         \min  I(X_{A t} \wedge Y_{A t} | \theta_1 = \tau_1) \   \qquad \qquad \text{ by } \eqref{eq:Conv_single_let_Markov} \\
 & =    \frac{1}{n} \sum \limits_{t=1}^n 
        \underset{
         \substack{ \mu_{X_{A t} Y_{A t} | \theta_1 = \tau_1} < \hspace*{-0.1cm} < \mu_{X_{A t}| \theta_1 = \tau_1}
         \times \mu_{ Y_{A t} | \theta_1 = \tau_1} \\ 
         \mathbbm{E}[d_{A \tau_1}(X_{A t}, Y_{A t}) |\theta_1 = \tau_1 ] 
          \leq \Delta_{\tau_1,t} - \Delta_{\min,A, \tau_1}}
         }
         \min  I(X_{A t} \wedge Y_{A t} | \theta_1 = \tau_1)   \ \ \quad \qquad 
         \text{ by Proposition } \ref{prop:prim_USRDf_equiv_form} \\  
 & = \frac{1}{n} \sum \limits_{t=1}^n \rho_A^{\cB}( \Delta_{\tau_1,t}, \tau_1 ) \\
 & \geq  \rho_{A}^{\cB} \Big ( \frac{1}{n} \sum \limits_{t=1}^n \Delta_{\tau_1, t}, \tau_1 \Big ) 
	\geq   \rho_A^{\cB} ( \Delta_{\tau_1}, \tau_1). \label{eq:Conv_Bay_eq1}
\end{align}
Now, \eqref{eq:Conv_Bay_eq1} holds for every $\tau_1 \in \Theta_1$, hence 
\begin{align} 
 R & \geq \underset{ \tau_1 \in \Theta_1 } \sup
     \rho_A^{\cB}( \Delta_{\tau_1}, \tau_1 ) \\
  & \geq \underset{ \{ \Delta_{\tau_1}, \ \tau_1 \in \Theta_1 \} \atop 
    \mathbbm{E}[\Delta_{\theta_1}] \leq \Delta } \inf \underset{ \tau_1 \in \Theta_1 } 
    \sup  \rho_A^{\cB}( \Delta_{\tau_1}, \tau_1 ) \label{eq:Conv_Bay_eq2} \\
  & = R_{A}(\Delta)
\end{align}
for $\Delta > \Delta_{\min,A}$.

\vspace*{0.2cm}

In the nonBayesian setting, the analog of Lemma \ref{l:single_letter_markov2}
is obtained similarly with $\theta = \tau, \ \theta_1 = \tau_1$ and 
\eqref{eq:l_s_block_markov}, \eqref{eq:l_claim}
replaced with appropriate conditional measures.
The proof of the converse is along the lines of the proof above, but with 
$\rho_A^{n \cB}(\Delta,\tau_1)$ in place of $\rho_{A}^{\cB}(\Delta_{\tau_1}, \tau_1)$,
and without the outer minimization with respect to $\{ \Delta_{\tau_1}, \tau_1 \in \Theta_1\}.$

\vspace*{0.2cm}

The converse proof for Theorem \ref{th:GMMS_SRDf} obtains immediately from the Bayesian
setting with the following changes:
$\Theta_1$ and $\Lambda(\tau_1), \ \tau_1 \in \Theta_1$, 
are taken to be singletons (rendering the infimum and supremum in \eqref{eq:Conv_Bay_eq2} superfluous) and 
\eqref{eq:ach_fnl_equiv} is used in place of Proposition \ref{prop:prim_USRDf_equiv_form}.

\qed

\vspace*{0.3cm}

\noindent The converse proof for Proposition \ref{prop:SRDf_GMF} involves an
approach which does not rely on Lemma \ref{l:single_letter_markov2} and is presented next.

\vspace*{0.2cm}

\noindent {\bf Converse proof for Proposition \ref{prop:SRDf_GMF}}:
Let $(f, \varphi)$ be an $n$-length $k$-FS code of code $R$ with 
$\mathbbm{E}[|| X_I^n -  \varphi(f(X_A^n)) ||^2]
\leq \Delta$. For $u \in I$ and $t = 1, \ldots,n,$ define
\begin{align}
\qquad {\widehat Y}_{ut } 
& = \mathbbm{E}[ X_{u t}| f(X_{A}^n) ] \\ 
& = \mathbbm{E}\big [ \mathbbm{E}[X_{u t}| X_{A}^n, f(X_{A}^n)] |  f(X_{A}^n) \big] \\ 
& = \mathbbm{E}\big [ \mathbbm{E}[X_{u t}| X_{A}^n] |  f(X_{A}^n) \big] 
\\ 
& = \mathbbm{E}\big [ \mathbbm{E}[X_{u t}| X_{A t}] |  f(X_{A}^n) \big] 
,  \quad 
\text{since } X_{A t}, X_{u t} \independent X_{A}^{n \setminus t}, X_{u}^{n \setminus t}\\
& = \mathbbm{E}[X_{\{u\} A}] {\bf \Sigma}_A^{-1} \mathbbm{E}[X_{A t}|f(X_{A}^n)] .
\end{align}
Notice that for $u \in I\setminus A$, 
$$ {\widehat Y}_{ut} = \mathbbm{E}[X_{\{u\} A}] {\bf \Sigma}_A^{-1} {\widehat Y}_{A t}, 
\qquad t = 1, \ldots, n.$$
By the optimality of the MMSE estimate 
\begin{align} \label{eq:mmse_form_gmf}
\Delta \geq 
\mathbbm{E}[ || X_{I}^n -  \varphi(f(X_A^n)) ||^2 ] \geq
\mathbbm{E}[ || X_{I}^n - {\widehat Y}_{I}^n ||^2 ] 
= \mathbbm{E}[  (X_{A}^n - {\widehat Y}_{A}^n )^T {\bf G}_{A, I} (X_A^n - {\widehat Y}_A^n)]
+ \Delta_{\min,A}.
\end{align}
The equality in \eqref{eq:mmse_form_gmf} can be seen to hold along the lines of
\eqref{eq:SRDf_ach_eq3}-\eqref{eq:SRDf_ach_eq4}.
Now, 
\begin{align}
\qquad  R & = \frac{1}{n} \log |f | 
  \geq  \frac{1}{n}H( f(X_{A}^n) ) \\ 
 & =  \frac{1}{n} I(X_A^n \wedge f(X_A^n) )\\
 & \geq \underset{f, \varphi \atop \mathbbm{E}[ || X_{I}^n - \varphi(f(X_A^n)) ||^2 ] \leq \Delta}
 \min  \frac{1}{n} I(X_A^n \wedge f(X_{A}^n) ) \\
 & \geq \underset{ \mu_{X_{A}^n Y_{A}^n} < \hspace*{-0.1cm} < 
 \mu_{X_{A}^n } \times \mu_{ Y_{A}^n}
 \atop  \mathbbm{E}[ (X_A^n - Y_A^n)^T {\bf G}_{A, I} (X_A^n - Y_A^n)] \leq \Delta  - \Delta_{\min,A} }
 \min  \frac{1}{n} I(X_A^n \wedge Y_{A}^n) \hspace*{5cm} \qquad \ \text{by } \eqref{eq:mmse_form_gmf}  \\
 & = \underset{ \mu_{X_{A}^n Y_{A}^n} < \hspace*{-0.1cm} < 
 \mu_{X_{A}^n } \times \mu_{ Y_{A}^n} \atop 
 \mathbbm{E}[ (X_A^n - Y_A^n)^T {\bf G}_{A, I} (X_A^n - Y_A^n)] \leq \Delta - \Delta_{\min,A} }
 \min \frac{1}{n} \sum \limits_{t=1}^n \Big ( I(X_{A t} \wedge X_{A}^{t-1}, Y_{A}^n  ) - 
 I(X_{A t} \wedge X_{A}^{t-1} )  \Big )  \\
 & = \underset{ \mu_{X_{A}^n Y_{A}^n} < \hspace*{-0.1cm} < 
 \mu_{X_{A}^n } \times \mu_{ Y_{A}^n} \atop 
 \mathbbm{E}[ (X_A^n - Y_A^n)^T {\bf G}_{A, I} (X_A^n - Y_A^n)] \leq \Delta  - \Delta_{\min,A} }
 \min \frac{1}{n} \sum \limits_{t=1}^n  I(X_{A t} \wedge X_{A}^{t-1}, Y_{A}^n  ) \hspace*{2.7cm} 
 \qquad \text{since } X_{A t} \independent X_{A}^{n \setminus t} \\
 & \geq \underset{ \mu_{X_{A}^n Y_{A}^n} < \hspace*{-0.1cm} < 
 \mu_{X_{A}^n } \times \mu_{ Y_{A}^n} \atop 
 \mathbbm{E}[ (X_A^n - Y_A^n)^T {\bf G}_{A, I} (X_A^n - Y_A^n)] \leq \Delta  - \Delta_{\min,A}}
 \min \frac{1}{n} \sum \limits_{t=1}^n I(X_{A t} \wedge  Y_{A t}  )  \\
 & \geq \underset{ \{ \Delta_t, \ 1 \leq t \leq n \} \atop 
 \frac{1}{n} \sum \limits_{t=1}^{n} \Delta_t \leq \Delta } \min \
 \frac{1}{n} \sum \limits_{t=1}^n 
 \underset{\mu_{X_{A t} Y_{A t}} < \hspace*{-0.1cm} < 
 \mu_{X_{A t} } \times \mu_{ Y_{A t}} \atop \mathbbm{E}[ (X_{A t} - Y_{A t})^T {\bf G}_{A, I}(X_{A t} - Y_{A t}) ]
 \leq \Delta_t  - \Delta_{\min,A}} \min \ I(X_{A t} \wedge  Y_{A t}  )  \label{eq:conv_GMF_abs_t} \\
 & = \underset{ \{ \Delta_t, \ 1 \leq t \leq n \} \atop 
 \frac{1}{n} \sum \limits_{t=1}^{n} \Delta_t \leq \Delta } \min \frac{1}{n} \sum \limits_{t=1}^n  \rho_{A}(\Delta_t) \\
 & = \rho_{A}(\Delta),
\end{align}
where \eqref{eq:conv_GMF_abs_t} is since
\begin{align} \label{eq:abs_cont_subset}
 \mu_{X_{A}^n Y_{A}^n}  < \hspace*{-0.1cm} < \mu_{X_{A}^n } \times \mu_{Y_{A}^n }
 \Rightarrow \mu_{X_{A t} Y_{A t}}  < \hspace*{-0.1cm} < \mu_{X_{A t} } \times \mu_{Y_{A t} }, \qquad t = 1, \ldots,n.
\end{align}
The claim \eqref{eq:abs_cont_subset} is easy to see by contradiction. 
Consider any real-valued rvs $Z_1, Z_2, Z_3$ with probability distribution
$\mu_{Z_1 Z_2 Z_3} < \hspace*{-0.1cm} < \mu_{Z_{1}} \times
\mu_{Z_{2}} \times \mu_{Z_{3}} $. Suppose, if possible,
$\mu_{Z_{1} Z_{2}}$ is not absolutely continuous with respect to $\mu_{Z_{1}} \times \mu_{Z_{2}}$, i.e.,
there exist $E_1, E_2 \in  {\cal B}(\mathbbm{R})$  such that 
\begin{align} \label{eq:abs_cont_claim}
 \mu_{Z_{1}}(E_1) \times \mu_{Z_{2}}(E_2) = 0 \quad \text{ and } \quad  \mu_{Z_1 Z_2}(E_1 \times E_2) \neq 0.
\end{align}
Considering a $E = E_1 \times E_2 \times  {\cal B}(\mathbbm{R}) $,
by \eqref{eq:abs_cont_claim} we have $(\mu_{Z_1 } \times \mu_{Z_2} \times \mu_{ Z_3})(E) 
=  \mu_{Z_1}(E_1) \times \mu_{Z_2}(E_2) \times \mu_{Z_3}(\mathbbm{R}) = 0$ but 
\begin{align}
 \mu_{Z_1 Z_2 Z_3}(E) \neq 0,
\end{align}
since $\mu_{Z_1 Z_2}(E_1 \times E_2) \neq 0, $ a contradiction, since 
$\mu_{Z_1 Z_2 Z_3} < \hspace*{-0.1cm} < \mu_{Z_{1}} \times
\mu_{Z_{2}} \times \mu_{Z_{3}} $.

\qed

Note that a converse proof for Theorem \ref{th:GMMS_SRDf} can be provided
along the lines of the converse proof for Proposition \ref{prop:SRDf_GMF}.
However, we prefer the current manner of presentation which provides for 
unity of ideas.

\noeqref{eq:SLM_block_markov}
\noeqref{eq:SLM_markov_part}

 \section{Acknowledgements}
 This work builds on previous joint works of the author with Prakash Narayan.
 The author also thanks Prakash Narayan for the many helpful discussions, 
 detailed review and comments of this manuscript.

\renewcommand\baselinestretch{0.9}
{\small
\providecommand{\bysame}{\leavevmode\hbox to3em{\hrulefill}\thinspace}

}

\appendix

\subsection{Proof of Lemma \ref{l:cont_condnl_dist}} \label{app:Bayesian}

Recall that the elements of the compact sets $\Theta$ and $\Theta_1$ are indexed
by $\tau$ and $\tau_1$, which take
values in $\mathbbm{R}^{m^2}$ and $\mathbbm{R}^{k^2}$ respectively. 
Now, every
$\tau \in \Theta$ can be seen as $\tau = (\tau_1, \tau_2)$ with $\tau_2$ taking values in
$\Theta_2$, a bounded subset of $\mathbbm{R}^{m^2 - k^2 }$. 
A continuous function over a compact set is uniformly continuous, hence,
for every $x_{\cM} \in \mathbbm{R}^m$, 
\begin{align}
 \nu_{X_{\cM}|\theta}(x_{\cM}| \tau_1, \tau_2) \ \text{ and } \ 
 \nu_{\theta}(\tau_1, \tau_2) 
\end{align}
are uniformly continuous in $(\tau_1, \tau_2)$.
Furthermore, as a function of $\tau_2$,  
$\nu_{X_{\cM}|\theta}(x_{\cM}| \tau_1, \tau_2)$ and $
 \nu_{\theta}(\tau_1, \tau_2) $ are bounded functions over bounded set $\Theta_2$
 and hence so is $\nu_{X_{\cM}|\theta}(x_{\cM}| \tau_1, \tau_2)
 \nu_{\theta}(\tau_1, \tau_2) $. By the Bounded Convergence Theorem, for 
 every $x_{\cM} \in \mathbbm{R}^{m}$ and $\tau_1 \in \Theta_1$
\begin{align}
 \underset{ {\tilde \tau}_1 \rightarrow \tau_1 } \lim \nu_{\theta_1}({\tilde \tau}_1)
 = 
 \underset{ {\tilde \tau}_1 \rightarrow \tau_1 } \lim \int \limits_{\Theta_2} 
 \nu_{\theta}({\tilde \tau}_1, \tau_2) \, d \tau_2
 = \int \limits_{\Theta_2} \nu_{\theta}(\tau_1, \tau_2) \, d \tau_2 = \nu_{\theta_1}(\tau_1)
 \label{eq:marginal_cvg}
\end{align}
and 
\begin{align} \label{eq:part_marginal_cvg}
\quad \underset{ {\tilde \tau}_1 \rightarrow \tau_1 } \lim \int \limits_{\Theta_2} 
 \nu_{X_{\cM}|\theta}(x_{\cM}|{\tilde \tau}_1, \tau_2) \nu_{\theta}(\tau_1, \tau_2) \, d \tau_2
 = \int \limits_{\Theta_2} \nu_{X_{\cM}|\theta}(x_{\cM}|{  \tau}_1, \tau_2) 
 \nu_{\theta}(\tau_1, \tau_2) \, d \tau_2, 
\end{align}
and thus from \eqref{eq:marginal_cvg} and \eqref{eq:part_marginal_cvg},
\begin{align}
 \underset{ {\tilde \tau}_1 \rightarrow \tau_1 } \lim 
 \nu_{X_{\cM}|\theta_1}(x_{\cM}|{\tilde \tau}_1)
 = 
 \nu_{X_{\cM}|\theta_1}(x_{\cM}|{  \tau}_1). 
\end{align}
Continuity of $\nu_{X_{\cM} | \theta_1}(x_{\cM}|\tau_1) $
in  $\tau_1$, implies that for 
$i=1, \ldots,m,$ and $   t=1, \ldots,n$,
\begin{align}
\qquad \mathbbm{E}\big [ \big (X_i - (\varphi(f(X_A^n)))_{i,t} \big )^2 | \theta_1 = \tau_1 ]  
\end{align}
is continuous in $\tau_1$. The continuity of 
\begin{align} \label{eq:cont_arg_tot_eq3}
 \mathbbm{E}\big [ || X_{\cM}^n - \varphi(f(X_A^n)) ||^2 \big |\theta_1 = \tau_1\big] 
\end{align}
in $\tau_1$ is now immediate. 
Since $\Theta_1$ is a compact set,
  \eqref{eq:cont_arg_tot_eq3} is  uniformly continuous in $\tau_1$.

\qed

\subsection{Proof of Lemma \ref{l:part_close}} \label{app:part_close}

\noindent First, observe that for every $\tau_1 \in \Theta_1, $
$\Lambda(\tau_1)$ is a convex, compact set. Now, the minimum in \eqref{eq:close_def_theta_1}
exists as that of a continuous function over a compact set. It is seen in a
standard manner that the convexity of $\Theta$ and $ \Theta_1$ imply the convexity of
\begin{align}
 g(\tau_{1,1}, \tau_{1,2}) \triangleq \underset{{\hat \tau} \in \Lambda(\tau_{1,2 } )} \min \
 \underset{{\check \tau} \in \Lambda(\tau_{1,1 } )} \min || {\hat \tau} - {\check \tau} ||
\end{align}
in $(\tau_{1,1}, \tau_{1,2} )$. Consequently, $g(\tau_{1,1}, \tau_{1,2})$ is continuous 
in $(\tau_{1,1}, \tau_{1,2})$. Define
\begin{align}
 D(\delta) \triangleq \underset{\tau_{1,1}, \tau_{1,2} \in \Theta_1 \atop  ||\tau_{1,1} - \tau_{1,2}|| 
 \leq \delta} \max \ \ g(\tau_{1,1}, \tau_{1,2}).
\end{align}
Clearly, $D(0) = 0$ 
and $D(\delta)$ is a continuous nondecreasing function of $\delta$ (Chapter 20, \cite{Ross}).

\vspace*{0.1cm}

Now, we prove the lemma by contradiction. Suppose if possible,  there exists an 
$\ep > 0$  such that for every $\delta > 0$ there exist 
$\tau_{1,1,\delta}, \tau_{1,2,\delta} \in \Theta_1$ with 
$|| \tau_{1,1, \delta} - \tau_{1,2, \delta} || \leq \delta$ and
\begin{align}
 g(\tau_{1,1, \delta}, \tau_{1,2, \delta}) > \ep.
\end{align}
Then,
\begin{align}
 0 = D(0) = \underset{\delta \rightarrow 0} \lim \ D(\delta)  
 = \underset{\delta \rightarrow 0} \lim \ \underset{\tau_{1,1, \delta}, \ \tau_{1,2, \delta}: \ 
 ||\tau_{1,1, \delta} - \tau_{1,2, \delta}|| 
 \leq \delta} \max \ \ g( \tau_{1,1, \delta}, \tau_{1,2, \delta} ) \geq \ep,
\end{align}
a contradiction. Hence, the lemma.

\vspace*{0.3cm}

\qed

\subsection{Proof of existence of the minimum and maximum in (\ref{eq:USRDf_Bayesian})}
\label{app:proof_min_max_Bay}

For every $\tau_1 \in \Theta_1,$
recall that $\rho_{A}^{\cB}(\delta, \tau_1)$ is, in effect, a rate distortion function,
hence its inverse $D_{A}^{\cB}(R,\tau_1)$ is well defined over $R \geq 0.$ 
Continuity of $\nu_{X_{\cM}|\theta_1}(x_{\cM}|\tau_1)$ in $\tau_1$ for every 
$x_{\cM} \in \mathbbm{R}^m$ implies the continuity of $D_{A}^{\cB}(R,\tau_1)$ 
in $\tau_1.$

\vspace*{0.1cm}

We now show the existence of the minimum and maximum on the right-side of 
\eqref{eq:USRDf_Bayesian}, i.e.,
\begin{align} \label{eq:USRDf_Bayesian2}
  \underset{ \{\Delta_{\tau_1}, \ \tau_1 \in \Theta_1 \} \atop 
 \mathbbm{E}[\Delta_{\theta_1}] \leq \Delta } \inf \underset{\tau_1 \in \Theta_1} \sup \ 
 \rho_A^{\cB}(\Delta_{\tau_1}, \tau_1)
 =  \underset{ \{\Delta_{\tau_1}, \ \tau_1 \in \Theta_1 \} \atop 
 \mathbbm{E}[\Delta_{\theta_1}] \leq \Delta } \min \underset{\tau_1 \in \Theta_1} \max \ 
 \rho_A^{\cB}(\Delta_{\tau_1}, \tau_1).
\end{align}
Denote the left-side of \eqref{eq:USRDf_Bayesian2} by $r$
and choose 
\begin{align}
 \Delta_{\tau_1}^{*}  = D_{A}^{\cB}(r, \tau_1), \  \quad \tau_1 \in \Theta_1.
\end{align}
The continuity of $D_{A}^{\cB}(r, \tau_1)$ in $\tau_1$ implies the 
continuity of $\Delta_{\tau_1}^{*}$ in $\tau_1$ and hence 
$\mathbbm{E}[\Delta_{\theta_1}^{*}]$ exists. A simple
proof of contradiction can be used to show that 
$\mathbbm{E}[\Delta_{\theta_1}^{*}] \leq \Delta.$ Thus, 
$\{ \Delta_{\tau_1}^{*}, \ \tau_1 \in \Theta_1 \}$ 
satisfies the constraint on the left-side of \eqref{eq:USRDf_Bayesian2}
and for every $\tau_1 \in \Theta_1,$
$\rho_{A}^{\cB}(\Delta_{\tau_1}^{*}, \tau_1) = r, \ $ with 
\begin{align}
 \underset{\tau_1 \in \Theta_1} \sup \rho_A^{\cB}(\Delta_{\tau_1}^{*}, \tau_1)
  = r
\end{align}
and hence \eqref{eq:USRDf_Bayesian2} holds.

\qed

\subsection{  Proof of  (\ref{eq:Bay_correct_est_dist}) }
\label{app:Bay_correct_est_dist}

Noting that ${\tilde \theta}_{1,n}(X_A^n)$ is a deterministic 
function of $X_A^n$, for $\tau_1 \in \Theta_1$ and $ {\tau}_{ _{1,1} } \in \Theta_{1, \delta}$
with $|| \tau_1 - {\tau}_{ _{1,1} }  || \leq 2 \delta $ and 
$P_{ {\tilde \theta}_{1,n} | \theta_1 }( {\tau}_{ _{1,1} } | \tau_1 ) > 0$,
\begin{align}
 \mathbbm{E} \big [ \big|\big| X_{\cM}^n - \varphi( {\tau}_{ _{1,1} } , 
  f_{ {\tau}_{ _{1,1} } }(X_A^n) ) &  \big|\big|^2 \big | 
  \theta_1 = \tau_1, {\tilde \theta}_{1,n} = {\tau}_{ _{1,1} }  \big ] \\
 \! & =  \! \frac{1}{ P_{ {\tilde \theta}_{1,n} | \theta_1 }( {\tau}_{ _{1,1} } | \tau_1 )}
  \mathbbm{E} \big [ \mathbbm{1} \big ( {\tilde \theta}_{1,n}(X_A^n) = {\tau}_{ _{1,1} }  \big ) 
  \big|\big| X_{\cM}^n -
  \varphi \big({\tau}_{ _{1,1} } , f_{{\tau}_{ _{1,1} } }(X_A^n )  \big ) \big|\big|^2 \big | 
  \theta_1 = \tau_1   \big ] \ \ \label{eq:Dist_Bay_USRDf_eq3}  \\
 \! & \leq \! \frac{1}{ P_{ {\tilde \theta}_{1,n} | \theta_1 }( {\tau}_{ _{1,1} } | \tau_1 )}
  \mathbbm{E} \big [  \big|\big| X_{\cM}^n -
  \varphi \big({\tau}_{ _{1,1} } , f_{{\tau}_{ _{1,1} } }(X_A^n )  \big ) \big|\big|^2 \big | 
  \theta_1 = \tau_1   \big ] \\
 \! & \leq \! \frac{1}{ P_{ {\tilde \theta}_{1,n} | \theta_1 }( {\tau}_{ _{1,1} } | \tau_1 )}
 \Big (  \mathbbm{E} \big [ \big|\big| X_{\cM}^n - 
  \varphi \big({\tau}_{ _{1,1} } , f_{{\tau}_{ _{1,1} } }(X_A^n )  \big ) \big|\big|^2 \big | 
  \theta_1 = {\tau}_{ _{1,1} } \big ] + \ep_1 \Big ) \ \ \text{by Lemma }
   \ref{l:cont_condnl_dist}   \\
 \! & \leq \! \frac{1}{ P_{ {\tilde \theta}_{1,n} | \theta_1 }( {\tau}_{ _{1,1} } | \tau_1 )}
     \big ( \Delta_{ {\tau}_{ _{1,1} }} + 2 \ep_1  \big ) \label{eq:Dist_Bay_USRDf_eq1} \\
  & \leq \! \frac{1}{ P_{ {\tilde \theta}_{1,n} | \theta_1 }( {\tau}_{ _{1,1} } | \tau_1 )}
     \big ( \Delta_{ {\tau}_{ {1} }} + 3 \ep_1  \big ) \label{eq:Dist_Bay_USRDf_eq2} 
\end{align}
where 

\noindent (i) \eqref{eq:Dist_Bay_USRDf_eq3} is since ${\tilde \theta}_{1,n}(X_A^n)$ is
a deterministic function of $X_A^n$;

\noindent (ii) it is seen along the lines of the achievability proof 
of Theorem \ref{th:GMMS_SRDf} that
 \begin{align}
  \mathbbm{E} \big [ || X_{\cM}^n - 
  \varphi ({\tau_{ _{ {1,1}}}}, f_{{\tau_{ _{ {1,1}}}}}(X_{A}^n)) ||^2
  \big | \theta_1 = {\tau_{ _{ {1,1}}}}  \big ] \leq \Delta_{{\tau_{ _{ {1,1}}}}}
  + \ep_1,
 \end{align}
and hence \eqref{eq:Dist_Bay_USRDf_eq1} is obtained;

\noindent (iii) $\Delta_{\tau_1}$ is continuous in $\tau_1$
over the compact set $\Theta_1$, hence, $\Delta_{\tau_1}$ is in fact uniformly continuous in $\tau_1$;
\eqref{eq:Dist_Bay_USRDf_eq2} now follows. % by the uniform continuity of the optimal 
%$\Delta_{ \tau_1}$ in $\tau_1$

\vspace*{0.2cm}

\noindent From \eqref{eq:Dist_Bay_USRDf_eq2}, the first term on the right-side of
\eqref{eq:Distortion_Bay_USRDf} is
\begin{align}
 \mathbbm{E}[ \mathbbm{1}( || {\tilde \theta}_{1,n} - \theta_1 || \leq 2 \delta ) 
  || X_{\cM}^n  -  \varphi(f(X_A^n)) ||^2 ]
  & \leq  \underset{ {\tilde \tau}_{ _{1}} \in \Theta_{1,\delta} } \sum  
      \mathbbm{E}[ \mathbbm{1}( ||  \theta_1 - {\tilde \tau}_{ _{1}} || \leq 2 \delta )
      (\Delta_{\theta_1} + 3 \ep_1 ) ] \\
  & \leq \mathbbm{E}[ \Delta_{\theta_1} ] + 3 \ep_1
 \qquad \qquad \text{by } \eqref{eq:excess_cover}  \\
  & \leq \Delta + 3 \ep_1.
\end{align}

\qed

\subsection{Proof of Lemma \ref{l:convexity_std}}
\label{app:convexity_std}

The right-sides of \eqref{eq:SRDf_GMMS} and \eqref{eq:SRDf-gauss-field}
are, in effect, the RDf for GMMS
with weighted MSE distortion criterion, and hence are
finite-valued, decreasing, convex, continuous functions of 
$\Delta > \Delta_{\min,A} $ and $\Delta  > \Delta_{\min,A, \tau_1}$,
respectively.

  \vspace*{0.1cm}
  
The right-sides of \eqref{eq:USRDf_Bayesian} and \eqref{eq:USRDf_nonBayesian} are 
clearly  nonincreasing functions of $\Delta$. Convexity of 
the right-sides of \eqref{eq:USRDf_Bayesian} and \eqref{eq:USRDf_nonBayesian}
follows from the convexity of $\rho_A^{\cB}(\delta, \tau_1)$
and $\rho_A^{n\cB}(\delta, \tau_1)$ using standard arguments;
continuity for $\Delta > \Delta_{\min,A,\tau_1}$ is a consequence. 
Finite-valuedness of \eqref{eq:USRDf_Bayesian} and  \eqref{eq:USRDf_nonBayesian} 
follows from the finite-valuedness of $\rho_{A}^{\cB}(\delta, \tau_1)$
and $\rho_{A}^{n\cB}(\delta, \tau_1)$ for $\delta > \Delta_{\min, A, \tau_1}$, respectively. 

\vspace*{0.2cm}

The convexity of the right-side of \eqref{eq:USRDf_Bayesian} can be shown explicitly as 
follows. 
Let $\tau_1(1)$ and $\tau_1(2)$ attain the maximum in \eqref{eq:USRDf_Bayesian}
at $\Delta = \Delta_1$ and $\Delta = \Delta_2$, respectively, where $\Delta_1 < \Delta_2$.
For $\Delta_1, \Delta_2 > \Delta_{\min,A}$, let
$\{ \Delta_{\tau_1}^1, \ \tau_1 \in \Theta_1 \}$ and $\{ \Delta_{\tau_1}^2, \ \tau_1 \in \Theta_1 \}$,
attain the minimum in \eqref{eq:USRDf_Bayesian}, respectively and are as in Appendix \ref{app:proof_min_max_Bay}.
For any $ 0 < \alpha < 1$,  and  $ {\tilde \tau}_{1} \in \Theta_1$,
\begin{align}
 \alpha R_{A}(\Delta_1) + (1-\alpha) R_{A}(\Delta_2) 
 & = \alpha \rho_{A}^{\cB}(\Delta_{{\tilde \tau}_{1}}^1,{\tilde \tau}_{1}) + (1-\alpha) \rho_{A}^{\cB}(\Delta_{{\tilde \tau}_{1}}^2, {\tilde \tau}_{1})  \\
 & \geq \rho_{A}^{\cB} ( \alpha  \Delta_{{\tilde \tau}_{1}}^1 + (1-\alpha) \Delta_{{\tilde \tau}_{1}}^2, {\tilde \tau}_{1} ) , \label{eq:l_convex_eq1} 
\end{align}
by the convexity of $\rho_{A}^{ \cB}(\delta, {\tilde \tau}_{1})$ in $\delta$.
Now, \eqref{eq:l_convex_eq1}
holds for every ${\tilde \tau}_{1} \in \Theta_1,$ hence 
\begin{align}
 \alpha R_{A}(\Delta_1) + (1-\alpha) R_{A}(\Delta_2) 
 &  \geq \underset{{\tilde \tau}_{1} \in \Theta_1} \sup \  \rho_{A}^{\cB} ( \alpha  \Delta_{{\tilde \tau}_{1}}^1 + (1-\alpha) \Delta_{{\tilde \tau}_{1}}^2, {\tilde \tau}_{1} ) \\
 & \geq  \underset{ \{\Delta_{\tau_1}, \tau_1 \in \Theta_1 \} \atop \mathbbm{E}[ \Delta_{\theta_1}] \leq \alpha \Delta_1 + (1-\alpha) \Delta_2 } \inf \underset{\tau_1 \in \Theta_1} \sup \ \rho_{A}^{\cB}(\Delta_{\tau_1}, \tau_1) \\
 & =  R_{A}( \alpha \Delta_1 + (1-\alpha) \Delta_2). 
\end{align}

\qed

\vspace*{2cm}

\vspace*{1.5cm}

\end{document}